\documentclass[aps,pra,superscriptaddress,showpacs,twocolumn,10pt,longbibliography,floatfix]{revtex4-1}
\usepackage{graphicx}
\usepackage{bm,amsmath,amsfonts}
\usepackage{amssymb}
\usepackage{amsbsy}
\usepackage{verbatim}
\usepackage{color}
\usepackage[]{units}
\usepackage{hyperref}
\usepackage[american]{babel}
\usepackage[utf8]{inputenc}
\usepackage[T1]{fontenc}
\usepackage{lmodern}
\begin{hyphenrules}{american}
\end{hyphenrules}

\begin{document}

\author{J. Stockhofe}
\email{jstockho@physnet.uni-hamburg.de}
\affiliation{Zentrum f\"ur Optische Quantentechnologien,
  Universit\"at Hamburg, Luruper Chaussee 149, 22761 Hamburg, Germany}

\author{P. Schmelcher}
\email{pschmelc@physnet.uni-hamburg.de}
\affiliation{Zentrum f\"ur Optische Quantentechnologien,
  Universit\"at Hamburg, Luruper Chaussee 149, 22761 Hamburg, Germany}
\affiliation{The Hamburg Centre for Ultrafast Imaging, Universit\"at Hamburg, Luruper Chaussee 149, 22761 Hamburg, Germany}

\title{Sub- and supercritical defect scattering \\in Schrödinger chains with higher-order hopping}

\begin{abstract}
We theoretically analyze a discrete Schrödinger chain with hopping to the first and second neighbors, as can be realized with zigzag arrangements of optical waveguides or lattice sites for cold atoms.
Already at moderate values, the second-neighbor hopping has a strong impact on the band structure, leading to the emergence of a new extremum located inside the band,
accompanied by a van Hove singularity in the density of states. The energy band is then divided into a subcritical regime with the usual unique correspondence between wave number and energy of the travelling waves,
and a supercritical regime, in which waves of different wave number are degenerate in energy. We study the consequences of these features in a scattering setup, introducing a defect that locally breaks 
the translational invariance. The notion of a local probability current is generalized beyond the nearest-neighbor approximation and bound states with energies outside the band are discussed.
At subcritical energies inside the band, an evanescent mode coexists with the travelling plane wave, giving rise to resonance phenomena in scattering. 
At weak coupling to the defect, we identify a prototypical Fano-Feshbach resonance of tunable shape
and provide analytical expressions for its profile parameters. At supercritical energies, we observe coupling of the degenerate travelling waves,
leading to an intricate wave packet fragmentation dynamics. The corresponding branching ratios are analyzed.
\end{abstract}

\maketitle
\section{Introduction}
In many fields of physics, understanding and controlling the propagation and scattering properties of waves is of profound importance \cite{Pike2002}.
Matter wave scattering from a localized defect within the framework of the Schrödinger equation
is one of the most basic, yet at the same time fundamentally important, problems of single particle quantum mechanics.
This is particularly true in view of present-day technology which makes it possible to fabricate devices
that are adequately described by basic quantum mechanical models, for example in electronics \cite{Ihn2004,Harrison2005}.
At the same time, ultracold atom experiments are also entering the regime of mesoscopic physics \cite{Brantut2012,Krinner2015,Chien2015},
promising to make the enormous toolbox for cold atomic gases accessible in quantum transport investigations.\\
The complexity of a scattering problem is not solely determined by the actual scatterer, but crucially depends on the structure of the asymptotic regions supporting the in- and outgoing waves.
For instance, the interplay of different transverse modes of quasi-one dimensional quantum waveguides gives rise to a rich variety of resonance phenomena,
even when considering only a point-like scatterer \cite{Bagwell1990,Kumar1991,Tekman1993,Noeckel1994,Kim1999a}, see also the confinement induced resonances in ultracold two-body collisions \cite{Olshanii1998,Dunjko2011}.
Nontrivial effects can also arise from an intricate discrete structure of the asymptotic leads, e.g. in carbon nanotubes or graphene \cite{Kostyrko1999,Kostyrko1999a,Rodrigues2013}.
Here, the detailed structure of the leads can often be modeled theoretically within a tight-binding approximation \cite{Goringe1997}, representing the material by an abstract multiply-connected lattice of discrete sites.\\
At the same time, tight-binding lattices beyond the standard first-neighbor approximation have also received considerable recent attention in the realm of cold atoms.
Protocols employing ions or Rydberg atoms have been proposed for the simulation of bosonic Hubbard models with long-range hopping \cite{Deng2008,Johanning2009,Olmos2013}.
A powerful method for enhancing specifically the second-neighbor hopping term in a Hubbard chain relies on rearranging the lattice sites in a zigzag geometry \cite{Efremidis2002}.
This has been implemented \cite{Dreisow2008} in evanescently coupled optical waveguide arrays \cite{Garanovich2012},
whose versatile applicability for the simulation of quantum scattering phenomena was also demonstrated in the realization of Klein tunneling \cite{Longhi2010b,Dreisow2012}.
Effects of the second-neighbor hopping on nonlinear excitations \cite{Efremidis2002,Kevrekidis2003,Szameit2009}, Bloch oscillations \cite{Wang2010,Dreisow2011,Stockhofe2015}, wave localization \cite{Golshani2013}, 
or the Mott insulator transition \cite{Sowinski2015} have been investigated.
Recently, several studies have explicitly addressed ultracold atoms in zigzag lattice geometries, both with interactions \cite{Greschner2013,Dhar2013} and without \cite{Metcalf2015}.
Here, the zigzag geometry could be obtained experimentally by singling out a strip from an extended two-dimensional triangular optical lattice \cite{Becker2010},
or using recently demonstrated techniques for designing essentially arbitrary in-plane optical potentials \cite{Henderson2009,Nogrette2014}.\\
The hallmark signature of sizable second-neighbor hopping in a noninteracting tight-binding chain is a deformation of the band structure,
causing the emergence of a split band edge \cite{Efremidis2002} (see also \cite{Sukhorukov2008}) and a corresponding van Hove singularity in the density of states located inside the band \cite{Koiller1981}.
We will investigate here the influence these features have when the zigzag lattice constitutes the asymptotic region of a scattering problem.
This setup has the advantage of providing nontrivial, yet generic, band structure features, 
while at the same time being accessible to analytical methods.
In fact, for similar reasons, it has been briefly suggested before as a minimal model 
for illustrating the Fano-Feshbach resonances \cite{Fano1961,Feshbach1958} in elastic light scattering off an obstacle \cite{Tribelsky2008}, but only a restricted parameter range of the model was considered there.
Beyond-first neighbor discrete Schrödinger lattice models also arise in the description of macromolecules such as DNA \cite{Zdravkovic2011},
and in this context the zigzag defect problem has been discussed in \cite{Koiller1981}, with a focus, however, on bound states instead of scattering properties.
Going beyond these works, we provide a comprehensive analysis of scattering from a localized defect in the zigzag geometry for a wide range of lattice parameters and energies,
identifying pronounced resonance features and the possibility of wave packet fragmentation induced by the second-neighbor hopping.
\\
Our presentation is structured as follows: In Sec. \ref{sec:setup}, we introduce the model and give a detailed discussion of the band structure features in the absence of the defect.
Sec. \ref{sec:current} explores the constraints imposed on stationary scattering solutions by continuity.
In Sec. \ref{sec:bound}, we discuss bound states at the defect, before entering the discussion of scattering states.
Here, we distinguish two energetic regimes, a subcritical one in which the interplay of a closed and an open channel induces a Fano-Feshbach resonance (Sec. \ref{sec:sub}),
and a supercritical one in which two open channels coexist and can be coupled to each other via scattering (Sec. \ref{sec:super}).
In both regimes we provide analytical results for the stationary scattering solutions and show corresponding simulations of the wave packet dynamics.
Finally, we summarize and conclude in Sec.~\ref{sec:conclusions} and give a brief outlook on future perspectives.

\section{Setup and band structure}
\label{sec:setup}
The discrete Schrödinger model we will be concerned with in the following has the general form
\begin{eqnarray}
 i \partial_\tau \psi_j = &&- \left(t_{1,j}^+ \psi_{j+1}+ t_{1,j}^- \psi_{j-1}\right) \nonumber \\
 &&- \left(t_{2,j}^+ \psi_{j+2}+t_{2,j}^- \psi_{j-2}\right)+V_j \psi_j.
\label{eq:eom}
\end{eqnarray}
Here, $\tau$ denotes time, $j\in \mathbb Z$ is the site index, and we allow for site-dependent first ($t_{1,j}^\pm$) and second ($t_{2,j}^\pm$) neighbor hopping and an on-site potential term $V_j$. The hopping matrix is constrained by $t_{1,j}^+=t_{1,j+1}^-$, $t_{2,j}^+=t_{2,j+2}^-$. We use dimensionless units throughout.
Such discrete Schrödinger models arise in a variety of contexts \cite{Kevrekidis2009}, with correspondingly different physical interpretations of the terms. 
When thinking of ultracold atoms, Eq.~(\ref{eq:eom}) describes the dynamics of a single particle (or a condensate of noninteracting bosons \cite{Trombettoni2001}) in a lattice potential within the lowest-band approximation,
i.e. taking into account a single localized Wannier mode per lattice site.
We will not consider interactions here, noting that  for many atomic species an effectively noninteracting limit can be prepared experimentally making use of internal-state Feshbach resonances \cite{Chin2010}.
Moreover, for the nearest-neighbor chain scattering results from the noninteracting system have been found to provide valuable information even in the presence of interactions \cite{Burioni2005,Burioni2006}.

\begin{figure}[ht]
\centering
\includegraphics[width=0.48\textwidth]{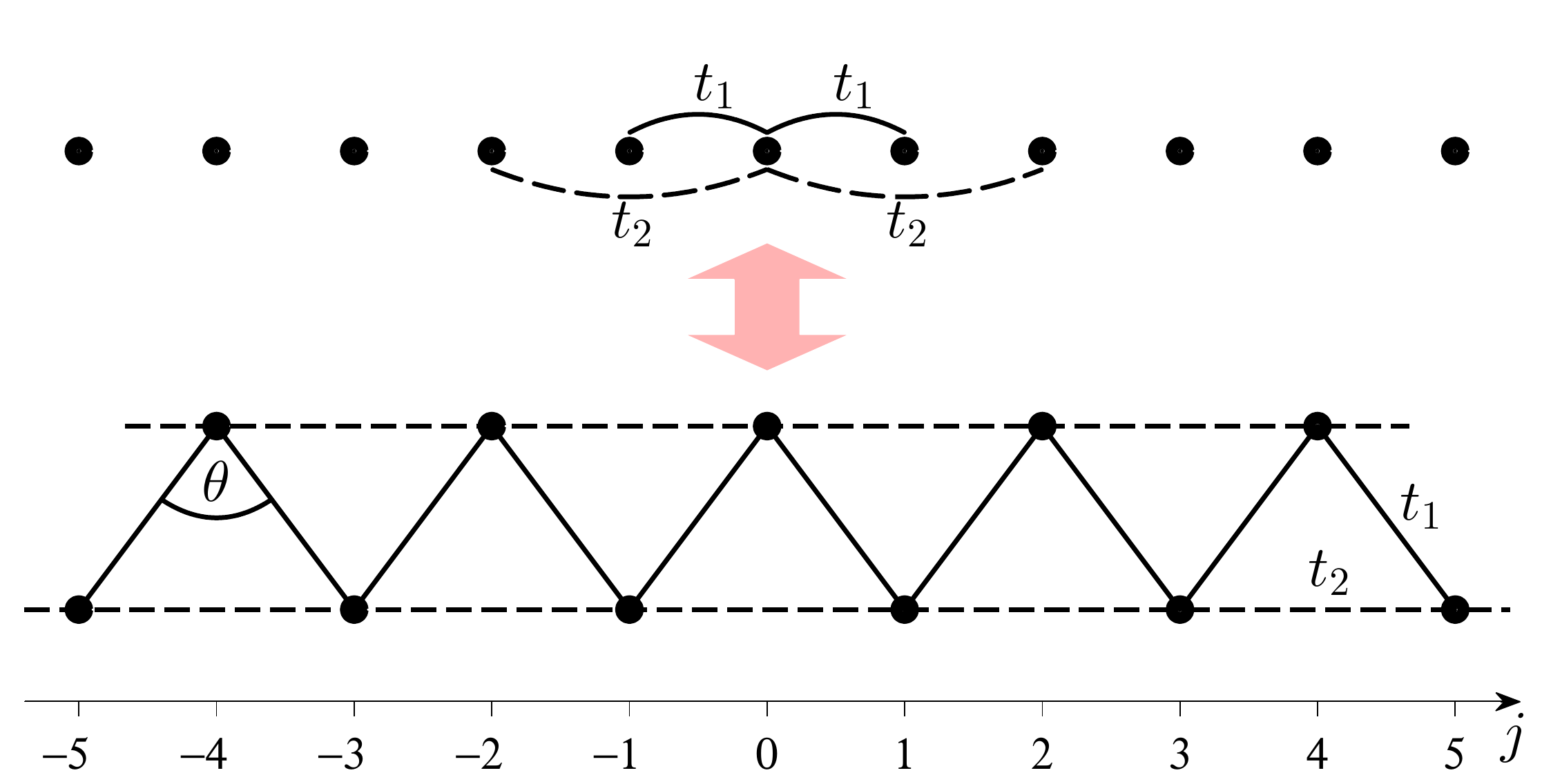}
\caption{(Color online) Sketch of the homogeneous discrete Schrödinger chain with hopping to the first and second neighbor and the zigzag arrangement of lattice sites which provides a realization of this model. 
Here, the opening angle $\theta$ tunes the relative strength of $t_2$ compared to $t_1$.\label{fig:sketchempty}}
\end{figure}

Let us first discuss the homogeneous system with $V_j = 0$, $t_{1,j}^\pm = t_1 >0$, $t_{2,j}^\pm = t_2>0$ for all $j$.
A sketch of this setup is shown in Fig.~\ref{fig:sketchempty}, illustrating also the connection to the zigzag arrangement of lattice sites that this model is tailored for. 
In the zigzag lattice, the opening angle $\theta$ effectively tunes the relative distances between first and second (in index) neighbor sites, and thereby also the relative values of the hopping amplitudes to the first and second neighbors \cite{Efremidis2002,Dreisow2008}. Hopping matrix elements beyond the second neighbor are neglected.

\begin{figure}[ht!]
\centering
\includegraphics[height=3.2cm]{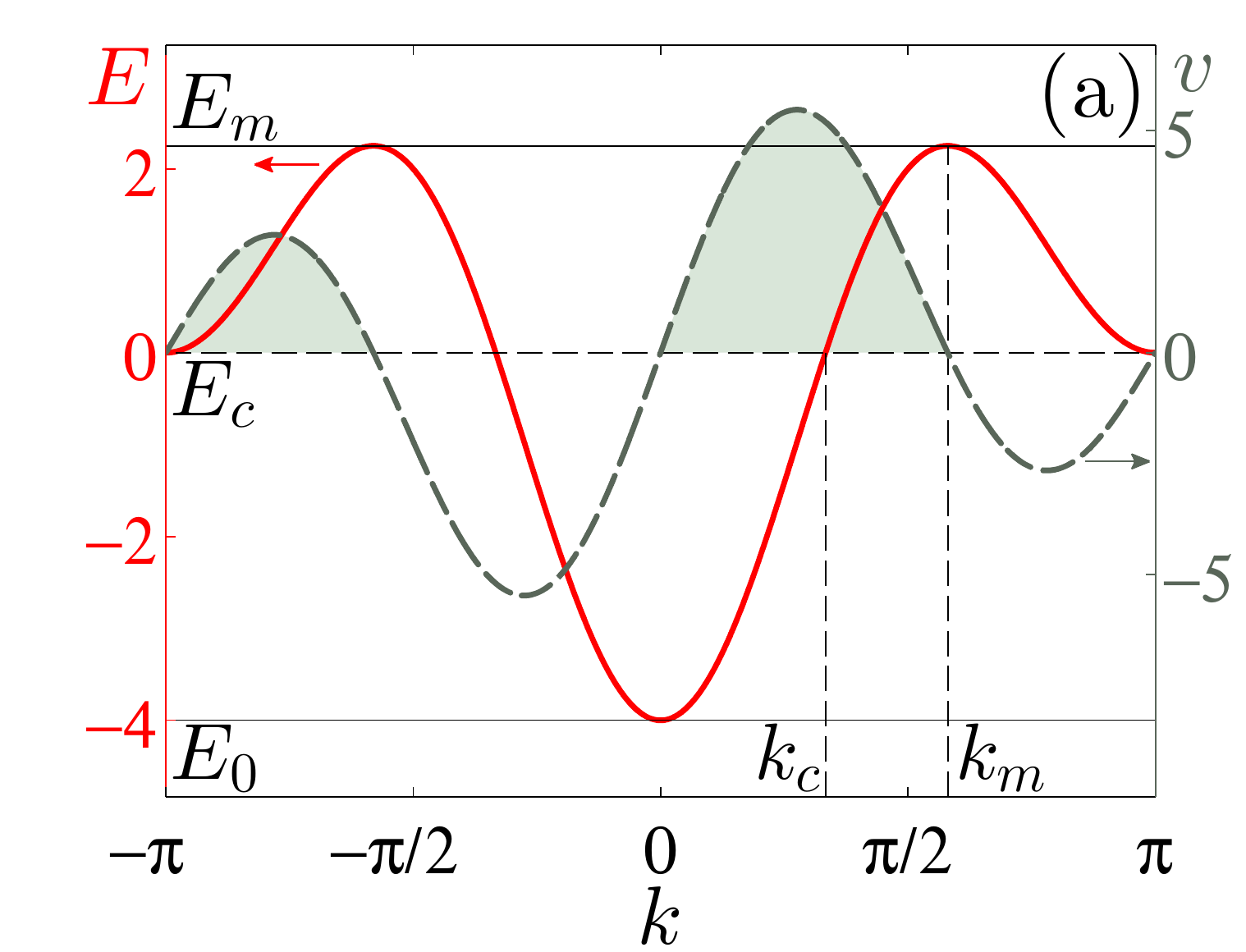}
\includegraphics[height=3.13cm]{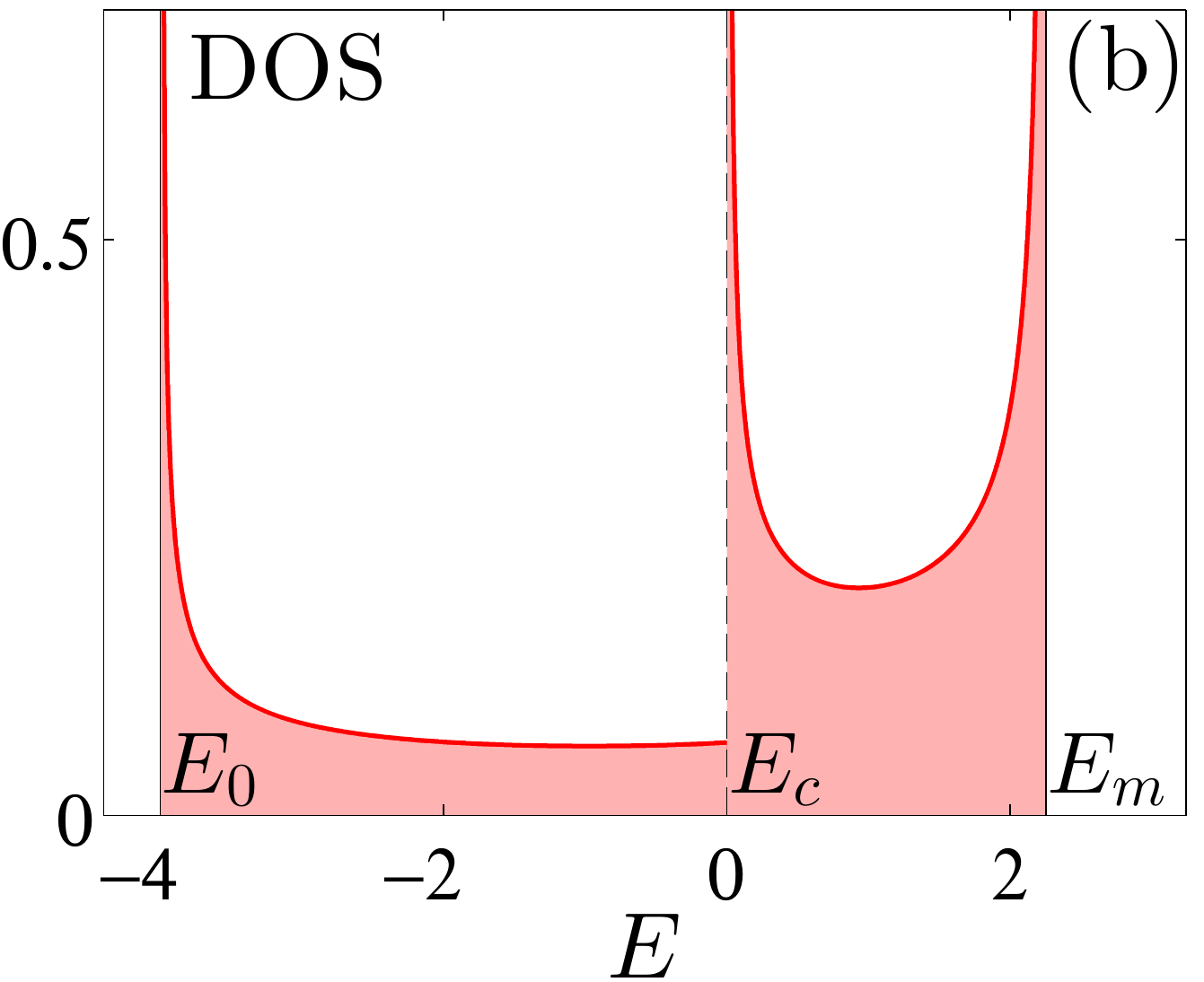}
\caption{(Color online) (a) Dispersion relation $E(k)$ (solid red line) and group velocity $v(k)$ (dashed grey line, shading highlights the intervals of positive $v$) and (b) normalized density of states for the homogeneous zigzag lattice model with $t_1=t_2=1$. \label{fig:BSDOS}}
\end{figure}

In the homogeneous case, the model is invariant under spatial translations and Eq.~(\ref{eq:eom}) admits stationary plane wave solutions
of the form $\psi_j \propto \exp \left[{i(kj-E(k) \tau)} \right]$,
where $k$ denotes the quasi-momentum which can be restricted to the first Brillouin zone $-\pi < k \leq \pi$.
The corresponding dispersion relation is given by 
\begin{equation}
 E(k)=-2t_1 \cos k - 2 t_2 \cos 2k.
\label{eq:dispersion}
\end{equation}
This dispersion curve is shown in Fig.~\ref{fig:BSDOS}(a) for the case $t_1=t_2=1$. 
Characteristically, when increasing the second-neighbor hopping $t_2$ the global minimum of $E(k)$ remains at $k=0$, $E_0 = E(0) = -2(t_1+t_2)$,
while the maximum of the dispersion is shifted away from $k=\pi$ (which itself transforms into a local minimum) as soon as $t_2 > t_1/4$, which is the case we will mostly focus on in the following.
In particular, this implies that above a certain energy $E_c$ within the band the dispersion is degenerate, 
in the sense that for $E > E_c$ there are two distinct quasi-momenta $k_1, k_2$
with $|k_1| \neq |k_2|$ satisfying $E(k_1)=E(k_2)$.
Remarkably, the group velocity
\begin{equation}
 v(k)=\frac{\text d E}{\text d k} = 2 t_1 \sin k + 4 t_2 \sin 2 k
\label{eq:v}
\end{equation}
is thus no longer in unique correspondence to the energy: The zigzag model permits wave packets of the same energy
that travel at different velocities in the lattice.
The group velocity curve $v(k)$ is also shown in Fig.~\ref{fig:BSDOS}(a).

\begin{figure}[ht]
\centering
\includegraphics[width=0.47\textwidth]{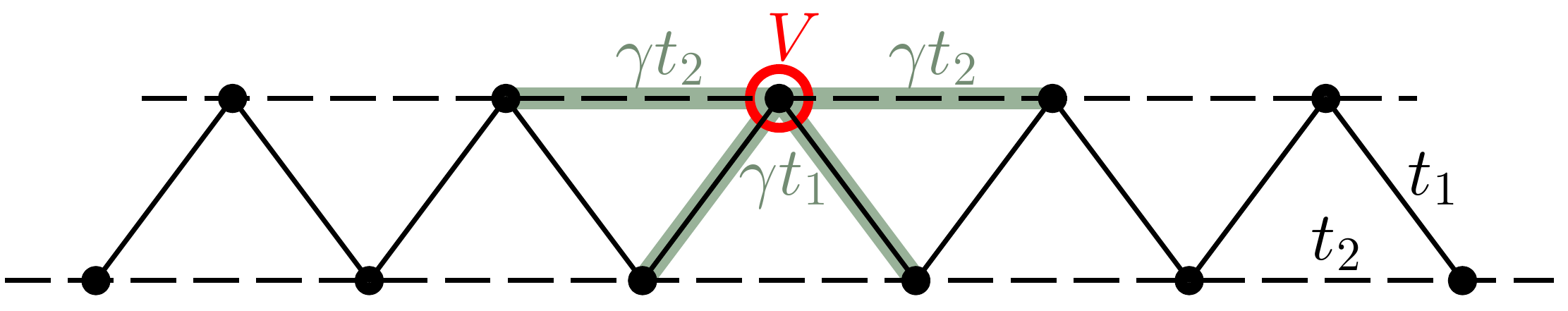}
\caption{(Color online) Sketch of the zigzag lattice containing a defect site with on-site potential $V$. Hopping to this defect site is modulated by a factor $\gamma$, compared to the asymptotic homogeneous hoppings. \label{fig:sketch}}
\end{figure}

For our model, the critical energy above which this travelling wave degeneracy occurs (given $t_2 > t_1/4$) explicitly reads as
\begin{equation*}
 E_c = E(\pm k_c)= 2(t_1-t_2), \quad k_c = \text{acos}\left(1 - \frac{t_1}{2 t_2}\right),
\end{equation*}
while the split upper band edge is located at 
\begin{equation*}
 E_m = E(\pm k_m)=2 t_2 + \frac{t_1^2}{4t_2}, \quad k_m = \text{acos}\left(-\frac{t_1}{4t_2} \right).
\end{equation*}
In the limit of dominant second-neighbor hopping, $t_2 \gg t_1$, this implies $k_c \rightarrow 0$, $k_m \rightarrow \pi/2$, such that
for large $t_2$ eventually the degeneracy region covers the whole band.
In contrast, for weak second-neighbor hopping, $t_2<t_1/4$, $E_c=2(t_1-t_2)$ marks the upper band edge at $k=\pm \pi$, while $E_m=2 t_2 + t_1^2/(4t_2)$ lies outside the band.
\\
Consider $t_2 > t_1/4$, then the qualitative deformation of the equienergy ``surface'' in $k$-space when crossing the critical energy $E_c$
may be thought of as the basic ingredient for observing a topological transition in the corresponding fermionic many-body system in the sense of \cite{Blanter1994}.
For a Fermi energy near $E_c$, the Fermi surface topology will be sensitive to small variations of the model parameters. 
Characteristically, the formation of additional stationary points of the dispersion curve at the critical energy $E_c$ 
leads to the emergence of a van Hove singularity in the density of states
which is shown in Fig.~\ref{fig:BSDOS}(b). \\ 
The above considerations already suggest that by tuning the hopping parameters
one can control the propagation and dispersion properties of wave packets in the homogeneous zigzag lattice, as has been suggested in \cite{Efremidis2002}.
Here, we analyze the effects of the band structure deformation induced by the second-neighbor hopping if the translational invariance of the system is broken 
and the plane waves are coupled to each other.
The most fundamental framework to study this is a scattering setup, with a localized defect that breaks the homogeneity, while the asymptotic zigzag regions are undisturbed. 
Specifically, we will consider a localized on-site potential, $V_j = V \delta_{j,0}$.
Away from this defect, in the asymptotic regions of the lattice, we again assume first- and second-neighbor hoppings $t_1>0$ and $t_2>0$ which are independent of the site index.
In general, a local manipulation of the lattice that causes the on-site potential shift at the defect site will also affect the hopping matrix elements to this site, cf. the discussion in \cite{Trompeter2003}.
To keep the number of parameters tractable, we will partially account for this by assuming that first- and second-neighbor hoppings that connect to the defect site are rescaled compared to the respective background values by a common factor $\gamma > 0$.
A local tuning of the hopping parameters only (without significantly altering the on-site potentials) is possible via a local variation of the inter-site distances \cite{Trompeter2003}. 
We will thus assume $\gamma$ and $V$ to be essentially independently tunable parameters. The scattering setup is sketched in Fig.~\ref{fig:sketch}.

\begin{figure}[ht]
\centering
\includegraphics[width=0.48\textwidth]{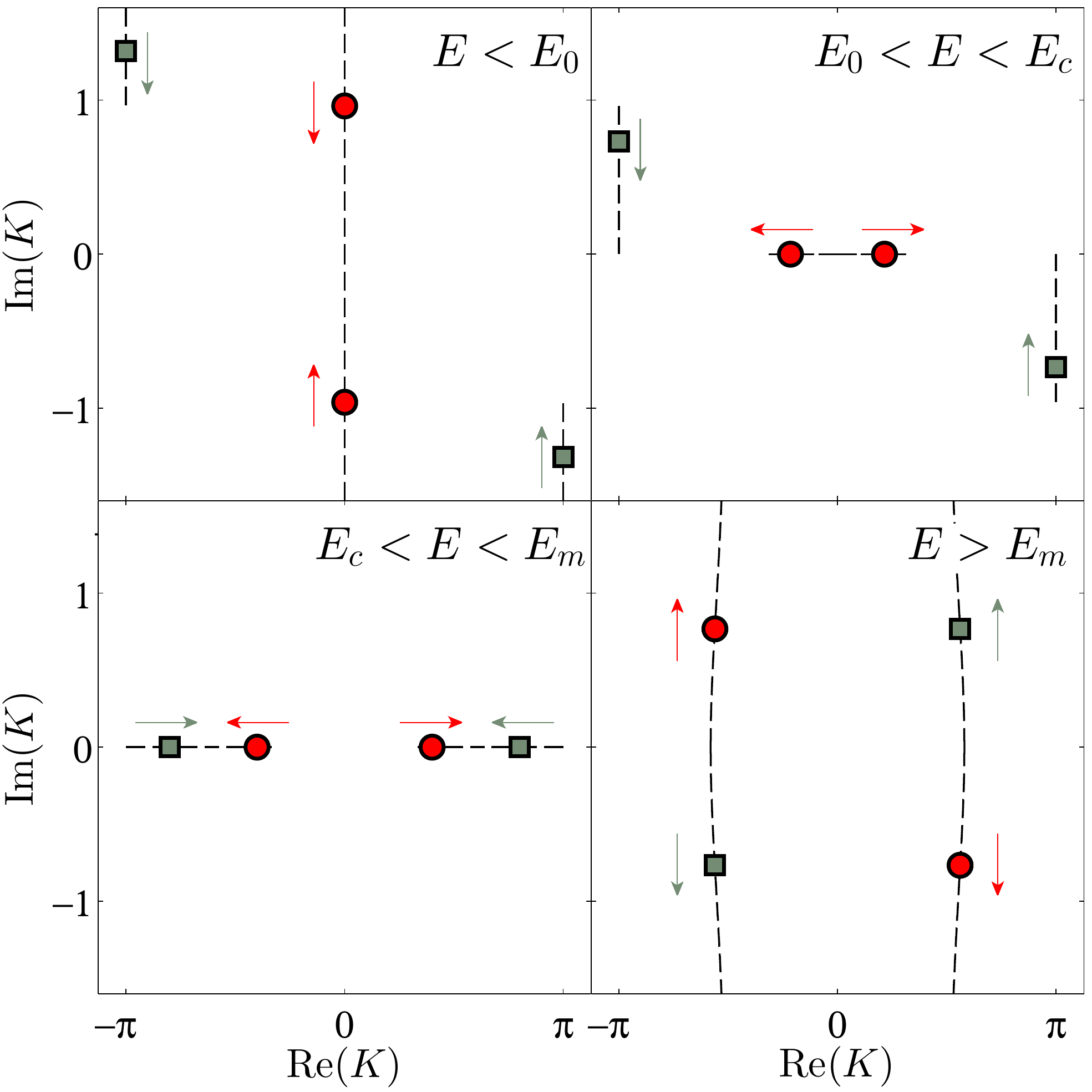}
\caption{(Color online) Complex solutions $K(E)$ according to Eq.~(\ref{eq:cosK}) for $t_1 =t_2 =1$ at different values of the energy $E$. Values of $K_1$ are indicated as red circles, values of $K_2$ by grey squares, while arrows indicate the direction in which these solutions move with increasing $E$. Qualitatively, the same picture is found for any $t_2 > t_1/4$. \label{fig:kbif}}
\end{figure}

Then, after the separation $\psi_j(\tau)=\phi_j \exp(-i E \tau)$ with a time-independent $\phi_j$ and the energy $E$, the stationary discrete Schrödinger system of equations explicitly reads
\begin{equation}
 E \phi_j = -t_1 \left( \phi_{j-1} + \phi_{j+1} \right) -t_2 \left( \phi_{j-2} + \phi_{j+2} \right)
 \label{eq:j2}
 \end{equation}
for $|j| > 2$, while
 \begin{eqnarray}
 E \phi_{\pm 2} &=& -t_1 \left( \phi_{\pm 1} + \phi_{\pm 3} \right) -t_2 \left( \gamma \phi_0 + \phi_{\pm 4} \right), \nonumber \\
 E \phi_{\pm 1} &=& -t_1 \left( \gamma \phi_0 + \phi_{\pm 2} \right) -t_2 \left( \phi_{\mp 1} + \phi_{\pm 3} \right), \nonumber \\
  (E-V) \phi_0 &=& -\gamma t_1 \left( \phi_{-1}+\phi_1 \right) -\gamma t_2 \left( \phi_{-2}+\phi_2 \right). \nonumber \\\label{eq:eomset}
\end{eqnarray}
Let us first discuss the asymptotic regions of $|j| > 2$, governed by Eq.~(\ref{eq:j2}).
Here, the homogeneous zigzag lattice discussed above is recovered. 
However, since we now look for piecewise solutions on the semiaxes $j>2$, $j<-2$ only, we cannot restrict to the travelling plane waves,
but need to take into account the possibility of evanescent waves which exponentially decay for $j \rightarrow \pm \infty$, respectively.
The fundamental solutions in the asymptotic regions are thus of the form $\phi_j \propto \exp (i K j)$ with a complex $K$. 
From Eq.~(\ref{eq:j2}), $K$ needs to satisfy the dispersion
relation $E=E(K)$ as in Eq.~(\ref{eq:dispersion}).
For our model, we can explicitly invert this to $K(E)$, with the result
\begin{eqnarray}
 \cos K_{1,2} (E) = -\frac{t_1}{4 t_2} \pm \sqrt{ \left( \frac{t_1}{4 t_2} \right)^2 - \frac{E}{4 t_2} + \frac{1}{2} },
\label{eq:cosK}
\end{eqnarray}
where for definiteness we choose the upper sign for $K_1$, the lower for $K_2$.
At a given energy $E$ the equation $E(K)=E$ thus has four complex solutions coming in pairs $\pm K_1, \pm K_2$.
These solutions, again for $t_1 = t_2 =1$, are visualized in Fig.~\ref{fig:kbif} at different values of $E$.
Qualitatively, the results do not change for other values of the hopping parameters as long as $t_2 > t_1/4$.
We can distinguish four different regimes.
For $E<E_0$, i.e. below the lower band edge, we find one pair of solutions that is purely imaginary, 
while the other has a non-vanishing imaginary part and a constant real part of $\pi$. 
Correspondingly, in this interval we have found no travelling waves.
At $E_0$, the former pair of solutions turns purely real: There is now one pair of real solutions (corresponding to the travelling plane waves found before)
and one pair of evanescent solutions of the form $\pm K_2 = \pi + i \kappa$ with $\kappa \in \mathbb R$.
At $E_c$, the imaginary part of this second pair also goes to zero, and for $E_c < E < E_m$ we are left with two real solutions,
corresponding to the two degenerate travelling waves we have observed in the band structure.
Finally, at $E_m$ there is a pairwise collision of the solutions in the complex plane
upon which they leave the real axis and form a complex quartet.

Let us contrast this to the case of weak second-neighbor hopping, $t_2 < t_1/4$. Then below $E_c$ (which in this case denotes the upper band edge) the situation is qualitatively the same as in panels (a) and (b) of Fig.~\ref{fig:kbif}, with the $K_1$ pair of solutions turning real at $E_0$ and then moving towards $\pm \pi$ with increasing energy. However, now at $E_c$ it is not the $K_2$ pair of solutions that also reaches the real axis, but instead the $K_1$ solutions reach $\pm \pi$ first and branch off along the imaginary axis, forming a second pair of the form $\pm(\pi + i \kappa)$ for $E_c < E < E_m$. Eventually, at $E_m$ there is again a pairwise collision (now taking place at the axes of $\pm \pi$ real part) upon which a complex quartet in the plane is formed. For $t_2 \rightarrow 0$ this second collision is pushed to infinite energies and the nearest-neighbor scenario is recovered.

From these considerations, one can already estimate the gross overall scattering features in the different energy ranges for $t_2 > t_1/4$:
Outside the band, below $E_0$ or above $E_m$, there are no travelling wave solutions at all, but bound states at the defect are expected.
In the subcritial energy regime, i.e. for $E_0 < E < E_c$, there is one travelling wave (open scattering channel) coexisting with an evanescent mode (closed channel), which may give rise to scattering resonances.
Finally, for supercritical energies with $E_c < E < E_m$, both channels are open and may be mixed during scattering events.
These three different regimes will be analyzed in detail in the following.\\

\section{Continuity equation}
\label{sec:current}
Before entering into the discussion of the scattering properties of the point defect, let us first discuss the constraints on the scattering coefficients due to local probability conservation.
Globally, the discrete Schrödinger equation (\ref{eq:eom}) preserves the norm $\sum_j |\psi_j|^2$. A local continuity equation is obtained by multiplying Eq. (\ref{eq:eom}) by $\psi_j^*$ and taking the imaginary part, which results in
\begin{eqnarray}
 \partial_\tau |\psi_j|^2 = -2 &&\text{Im}( t_{1,j}^+ \psi_j^* \psi_{j+1}+ t_{2,j}^+ \psi_j^* \psi_{j+2}) \nonumber\\
  +2 &&\text{Im} (t_{1,j}^- \psi_j \psi_{j-1}^*+ t_{2,j}^- \psi_j \psi_{j-2}^*).
\label{eq:continuity}
\end{eqnarray}
This identity holds at each $j$ and independently of the set of on-site potentials $\{V_j \}$.
It may be thought of as a Kirchhoff-type balance equation, relating the change of probability at a site to the currents in the four links connected to this site.
Now, in spite of the multi-connectedness of the lattice due to the second-neighbor hopping, we can rewrite this in the form of a standard one-dimensional continuity equation by defining a local current as
\begin{equation}
 J_j = 2 \sum_{\alpha=1}^2 \sum_{\beta=0}^{\alpha-1} t_{\alpha,j-\beta}^+ \text{Im}\left(\psi^*_{j-\beta}\psi_{j+\alpha-\beta} \right).
\end{equation}
Then Eq.~(\ref{eq:continuity}) reads as
\begin{equation}
 \partial_\tau |\psi_j|^2 + J_j - J_{j-1} = 0,
\label{eq:continuity2}
\end{equation}
where the symmetry $t_{\alpha,j}^+=t_{\alpha,j+\alpha}^-$ of the hopping matrix has been used. In particular, Eq.~(\ref{eq:continuity2}) implies that for a stationary solution the current is a global constant, $J_j \equiv J$ for all $j$.
In scattering scenarios, this equation relates the coefficients of the fundamental solutions in the asymptotic regions left and right of the defect \cite{Sautet1988}, see below.
We note that a generalization of the above concepts, in particular of the expression for the local current, for one-dimensional lattices with more extended hopping (beyond the second neighbor) is possible.
\section{Bound states}
\label{sec:bound}
\begin{figure}[ht]
\centering
\includegraphics[width=.48\textwidth]{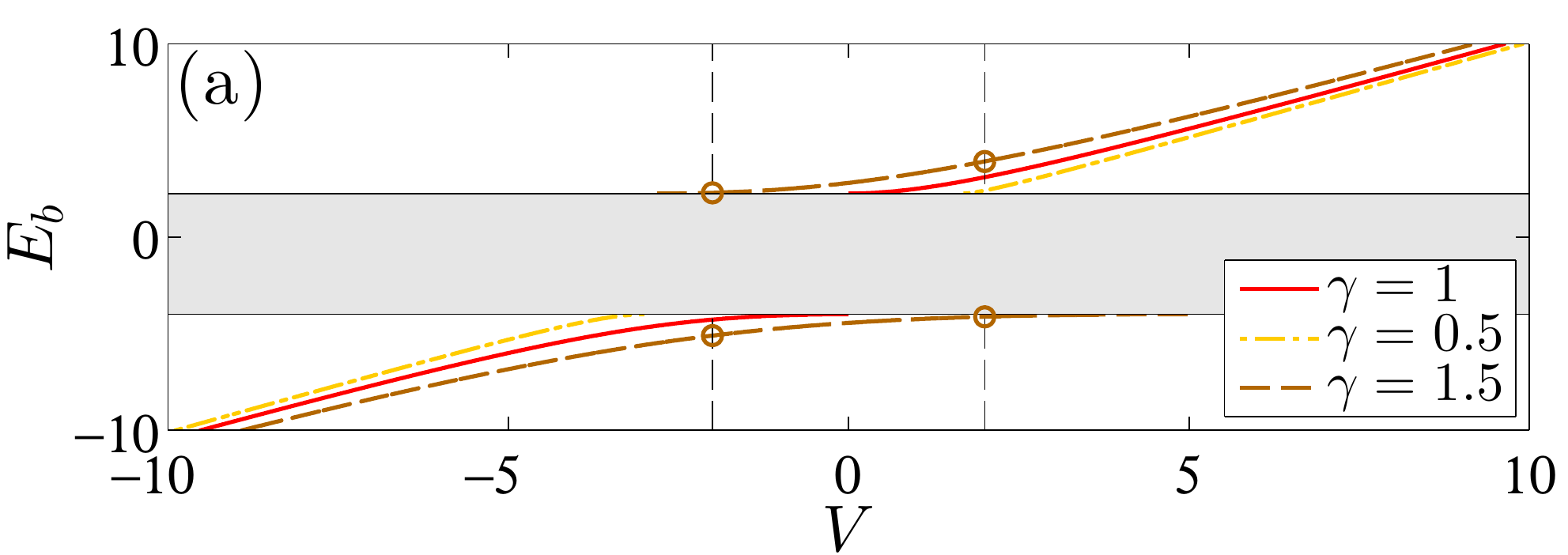} \\
\includegraphics[width=.235\textwidth]{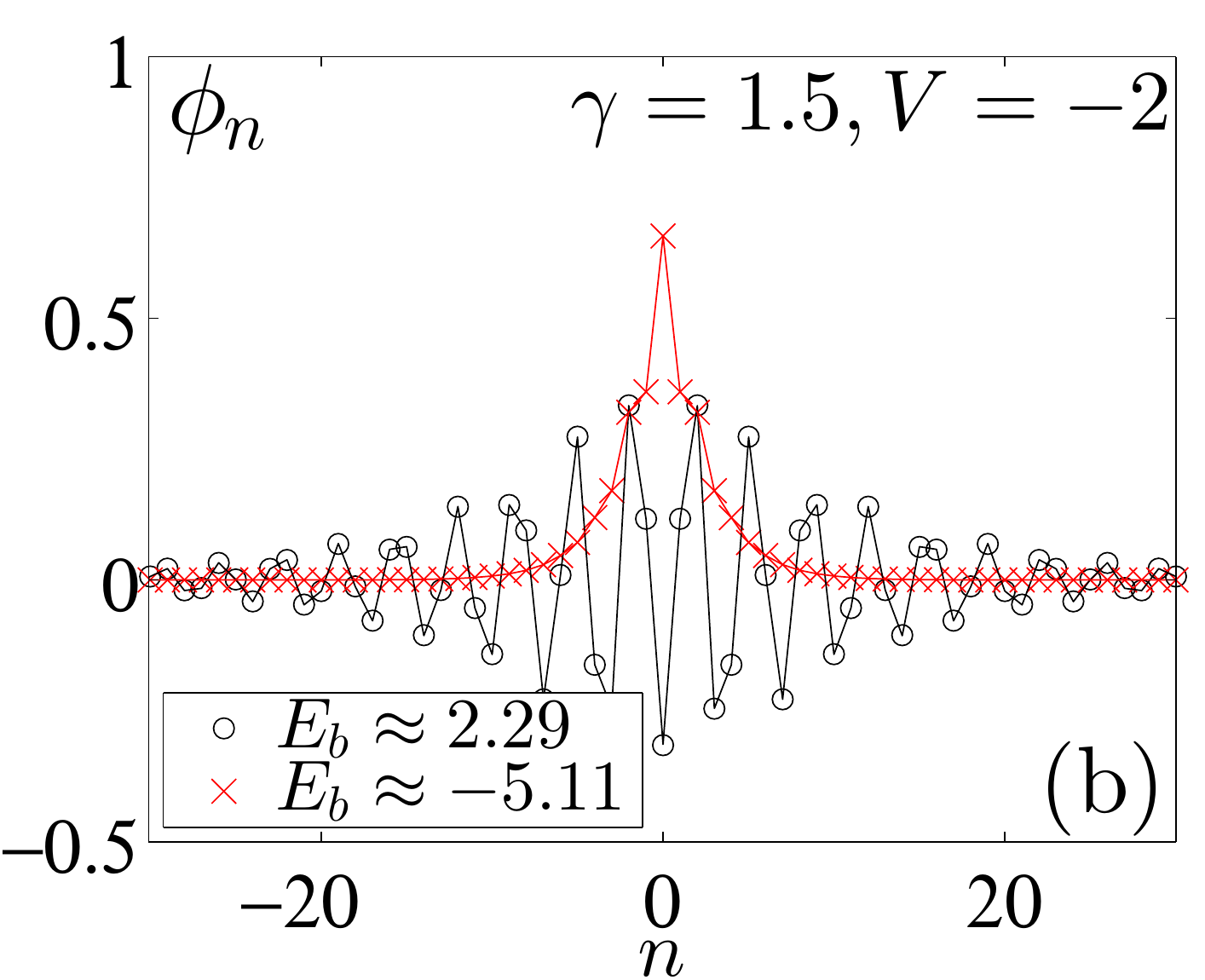} 
\includegraphics[width=.235\textwidth]{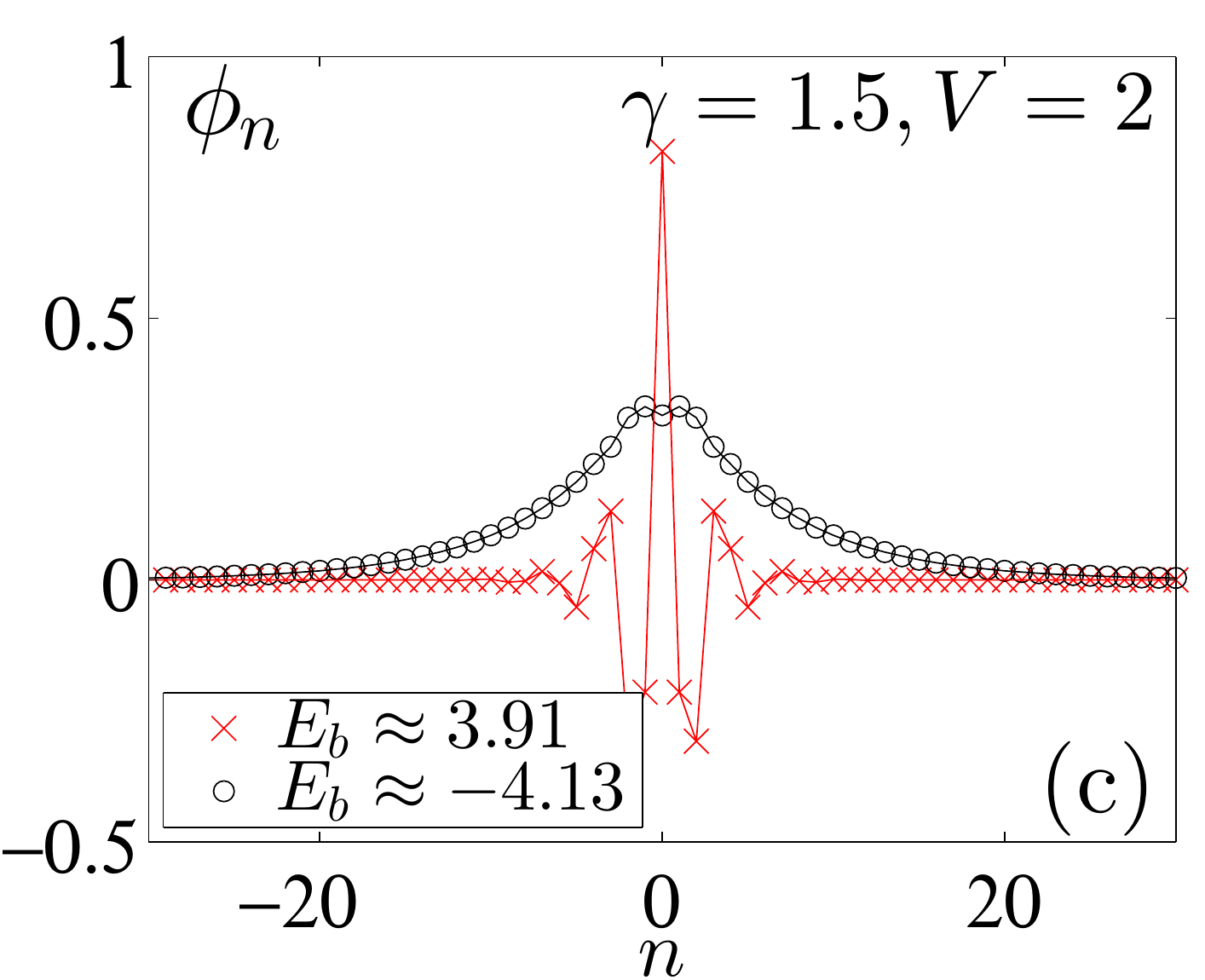} 
\caption{(Color online) (a) Energy of bound state(s) vs. defect potential for varying coupling strength $\gamma$ and $t_1=t_2=1$. The energy band in which propagating solutions exist is indicated in grey. (b,c) Profiles of the two bound states (above and below the band) at $V=\pm 2$, $\gamma=1.5$, as marked by brown circles and dashed vertical lines in (a). \label{fig:bound}}
\end{figure}

We now turn to the zigzag scattering setup as shown in Fig.~\ref{fig:sketch}. In this section we briefly discuss bound states at the defect,
having energies $E$ outside the band.
Then, as seen above, all solutions of $E=E(K)$ have a nonvanishing imaginary part.
Let $K_1, K_2$ denote the two solutions whose imaginary part is positive (to prevent asymptotic exponential growth), then we search for bound states in the form
 \begin{eqnarray*}
  \phi_j &=& A_1 e^{-i K_1 j} + A_2 e^{-i K_2 j}, \qquad j < 0,\\
  \phi_j &=& B_1 e^{i K_1 j} + B_2 e^{i K_2 j}, \qquad \quad j >0,
\end{eqnarray*}
with complex coefficients $A_1, A_2, B_1, B_2$.
Inserting this into Eqs.~(\ref{eq:eomset}) yields a homogeneous system of linear equations for $(A_1, A_2, B_1, B_2, \phi_0)$ which has nonzero solutions only if
the coefficients satisfy
\begin{eqnarray}
 B_1 &=& A_1, \, B_2 = A_2 = -\frac{\sin K_1}{\sin K_2} A_1, \label{eq:bounda}\\
 \gamma \phi_0 &=& \left(1  -\frac{\sin K_1}{\sin K_2} \right)A_1. \label{eq:boundb}
\end{eqnarray}
and simultaneously
\begin{eqnarray}
 V &=& E(1-\gamma^2) \nonumber \\&&+ \frac{2 i \gamma^2 \sin K_1 \sin K_2}{\sin K_2 - \sin K_1} \sqrt{t_1^2 -4t_2 E + 8 t_2^2}. 
\label{eq:bound}
\end{eqnarray}
Both below and above the band
it is readily found that the expression on the right hand side of Eq.~(\ref{eq:bound}) is real.
Thus, one can conclude that for any bound state energy $E_b$ outside the band and any fixed $\gamma$, there is precisely one value of the defect potential $V$ that produces precisely one bound state at this energy.
This $V(E_b, \gamma)$ is given by Eq.~(\ref{eq:bound}). Fig.~\ref{fig:bound}(a) shows the corresponding dependence of $V$ (horizontal axis) on the bound state energy $E_b$ (vertical axis) for different values of $\gamma$.
Conversely, it illustrates that for $\gamma=1$ each defect potential $V \neq 0$ also produces precisely one bound state (which lies below the band for $V<0$ and above the band for $V>0$).
In contrast, if $\gamma <1$ then for $V$ in a region around zero there is no bound state at all, while if $\gamma > 1$, for $V$ in a region around zero there are two bound states (one above, one below the band).
For large values of the defect potential $|V|$, the bound state energy $E_b \approx V$, irrespectively of $\gamma$, as is expected from perturbation theory.
This limit is accompanied by an increasing localization of the eigenmode when energetically moving away from the band.
Obviously, also in the decoupling limit $\gamma \rightarrow 0$ we find $E_b \rightarrow V$ and asymptotically perfect localization at the defect.
The relations between the coefficients in Eqs.~(\ref{eq:bounda},\ref{eq:boundb}) ensure that the global phase of the wave function can be chosen such that it is real (as is to be expected from time-reversal symmetry).
Thus, the continuity equation (\ref{eq:continuity2}) is trivially satisfied with $J = 0$.
Qualitatively, these features have been observed before for models with nearest-neighbor hopping only \cite{Trompeter2003} and for weak second-neighbor hopping \cite{Koiller1981}.
In the limit of $t_2 \rightarrow 0$, the imaginary part of $K_2$ diverges, such that $\sin K_1 / \sin K_2 \rightarrow 0$.
Then Eq.~(\ref{eq:bound}) reduces to $V=-2t_1 \cos K_1 + 2 t_1 \gamma^2 \exp(i K_1)$, from which it follows that $\exp(-i K_1) = -V/(2t_1) \pm \sqrt{V^2/(2t_1)^2+2 \gamma^2-1}$,
recovering the nearest-neighbour result of \cite{Trompeter2003}, while at the same time Eqs.~(\ref{eq:bounda},\ref{eq:boundb}) imply that $A_2, B_2 \rightarrow 0$, $\gamma \phi_0 \rightarrow A_1$.

A crucial difference to the nearest-neighbor model, however, is the absence of symmetry between bound states above and below the band.
When $t_2=0$, the staggering transformation $\phi_j \rightarrow (-1)^j \phi_j$ maps a stationary solution at energy $E$ for the potential strength $V$ to 
a solution at energy $-E$ for the potential strength $-V$. 
In particular, this implies that the bound state(s) at $+V$ and $-V$ share the same density profiles and differ only in their local phases, cf. \cite{Trompeter2003}.
Invariance under the staggering transformation does not hold in the presence of the second-neighbor hopping term. 
Thus, upon changing the sign of the defect potential $V$, the corresponding bound states feature different density profiles.
This can be seen in panels (b) and (c) of Fig.~\ref{fig:bound}.
Notably, interference between the two evanescent waves constituting the bound state allows for a nonmonotonic density decay away from the defect site.
This feature is most apparent in the extreme limit of $t_2 \gg t_1$, where the zigzag lattice separates into two essentially decoupled chains and thus the bound states at the defect will dominantly populate every second site.

\section{Scattering at subcritical energies}
\label{sec:sub}
\begin{figure}[ht]
\centering
\includegraphics[width=0.23\textwidth]{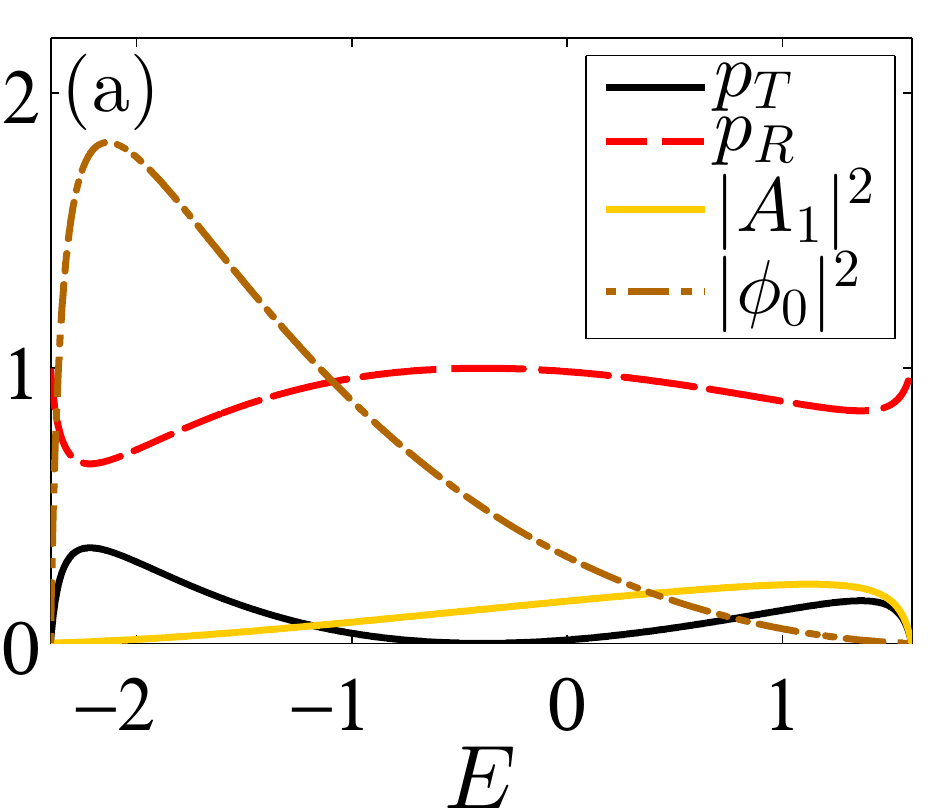} 
\includegraphics[width=0.23\textwidth]{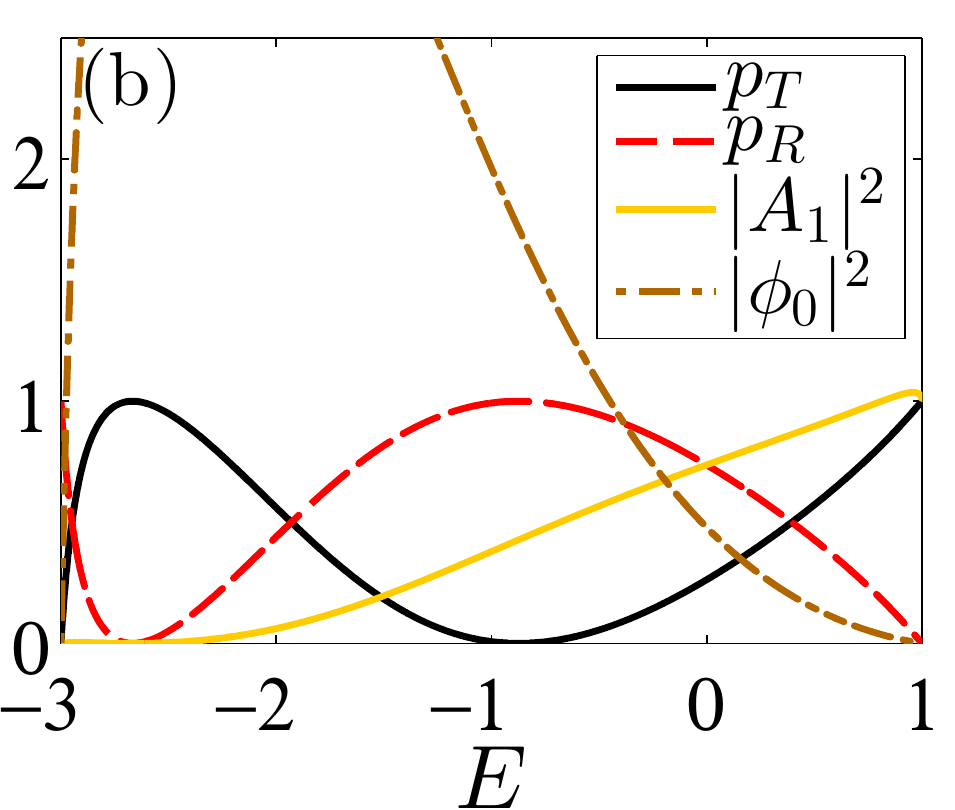} 
\caption{(Color online) Parameters of stationary scattering solutions in the subcritical energy range $E_0 < E < E_c$: Probability of transmission/reflection, squared amplitude of the evanescent mode and occupation of the defect site. 
Model parameters are chosen as $t_1 = 1$, $\gamma=0.5$, $V=-2$, and (a) $t_2 = 0.2$, (b) $t_2 = 0.5$.
\label{fig:singlek}}
\end{figure}

From here on we focus on energies lying inside the band of propagating states.
In this section we discuss scattering off the defect at subcritical energies $E$, satisfying $E_0 < E < E_c$.
For weak second-neighbor hopping, $t_2 < t_1/4$, this energetic regime covers the whole energy band ($E_c$ coincides with the upper band edge), while for $t_2 > t_1/4$ the critical $E_c$ lies inside the band.
The following discussion applies to both cases likewise.\\ 
For $E_0 < E < E_c$, there is a unique real wave number $k>0$ with $E(k)=E$, corresponding to $K_1 > 0$ as given by Eq.~(\ref{eq:cosK}).
Apart from this, there is a pair of staggered evanescent modes with wave numbers $K_2 = \pi \pm i \kappa$ where $\kappa > 0$ and $E(K_2)=E$.
Generally, the interplay of an open and a closed scattering channel is expected to lead to resonance effects,
as have been observed for instance in numerous variations of tight-binding lattices with side-coupled defects \cite{Miroshnichenko2010}.
In the following, we will explicitly work out the general transmission properties of the zigzag-defect model and in a second step specialize the results to the weak-coupling resonance regime.\\
We first note that $\cos K_1 + \cos K_2 = -t_1/(2 t_2)$ according to Eq.~(\ref{eq:cosK}), so the parameters of the travelling and the evanescent wave, respectively, at the same energy are related through
\begin{equation*}
 \kappa = \text{acosh} \left(\frac{t_1}{2t_2} + \cos k \right),
\end{equation*}
where $k<k_c$ ensures that $\kappa$ is real. To obtain the stationary scattering solutions, we employ the Ansatz:
\begin{eqnarray*}
 \phi_j &=& e^{i k j} + R e^{-i k j} + A_1 e^{-i j(\pi + i \kappa)}, \quad j <0, \\
 \phi_j &=& T e^{i k j} + A_2 e^{i j(\pi + i \kappa)}, \qquad \qquad \quad j >0,
\end{eqnarray*}
where $R, T, A_1, A_2$ and $\phi_0$ are unknown complex numbers and the signs have been chosen such that there is no asymptotic exponential growth.
Inserting this into Eqs.~(\ref{eq:eomset}) now results in a $5 \times 5$ inhomogeneous linear system of equations which can be solved for $(R, T, A_1, A_2, \phi_0)$ analytically, yielding
\begin{eqnarray}
 R &=& \left[i \left( \frac{\sin k}{\sinh \kappa} + \frac{\gamma^2 v(k)}{V+E(\gamma^2-1)} \right) -1 \right]^{-1} \label{eq:R}, \\
 T &=& R+1 \nonumber \\
&=& \left[1+i \left( \frac{\sin k}{\sinh \kappa} +  \frac{\gamma^2 v(k)}{V+E(\gamma^2-1)} \right)^{-1} \right]^{-1},  \label{eq:T} \\ 
 A_1 &=& A_2 = -i \frac{\sin k}{\sinh \kappa} R \label{eq:A1}, \\
\phi_0 &=& (R + A_1 + 1)/\gamma \label{eq:phi0}.
\end{eqnarray}
Here, $v(k)$ denotes the group velocity at $k$ as given by Eq.~(\ref{eq:v}).
In the limit of $t_2 \rightarrow 0$ (such that $\sinh \kappa \rightarrow \infty$), we recover the result of \cite{Sautet1988} for a nearest-neighbor chain disturbed by 
a defect.
Taking also $\gamma=1$, the above expressions reduce to those for a potential-only (no modulation of the hopping) impurity in a nearest-neighbor chain,
$R =\left[2i t_1 \sin k/V -1 \right]^{-1}$, $T=\left[1+ i V/(2t_1 \sin k) \right]^{-1}$, $A_1=A_2=0$.\\
Using the continuity equation (\ref{eq:continuity2}), we find that the contributions of the evanescent modes to the current cancel out and we are left 
with the probability conservation relation  $|R|^2 + |T|^2 = 1$, where the probability for reflection is explicitly given by
\begin{equation}
 p_R =|R|^2 = \left[ 1+ \left( \frac{\sin k}{\sinh \kappa} + \frac{\gamma^2 v(k)}{V+E(\gamma^2-1)} \right)^2 \right]^{-1}.
\end{equation}
We can immediately read off a number of features here.
First, if $V=0$ and $\gamma=1$ we recover the trivial case of the homogeneous lattice with $p_R = 0$ at all energies (we will exclude this case in the following).
Further, if $\gamma = 0$ (i.e., the defect site is fully detached from the lattice), the transmission and reflection probabilities are independent of $V$, as expected, and interestingly only
depend on the ratio $\sin k /  \sinh \kappa$.
For arbitrary $V$, $\gamma$, in the low energy limit of $k\rightarrow 0$ we find $p_R \rightarrow 1$, so the transmission probability $p_T = 1-p_R$ vanishes, and neither the defect site 
nor the evanescent mode are notably excited, $\phi_0 \rightarrow 0$, $A_1 \rightarrow 0$, as is seen from Eqs.~(\ref{eq:A1},\ref{eq:phi0}).
In contrast, approaching the upper edge of the energy interval, $E \rightarrow E_c$, we need
to distinguish two cases.
For $t_2 < t_1/4$, $E_c$ coincides with the upper band edge and the corresponding $k=\pi$, while $\kappa$ remains finite. Thus, $p_R \rightarrow 1$ and the transmission probability drops to zero when approaching $E_c$.
In contrast, for $t_2 > t_1/4$ we have $\sinh \kappa \rightarrow 0$ when $E$ approaches $E_c$ from below, while $\sin k_c$ and $v(k_c)$ remain finite and thus the transmission probability becomes unity. 
This full transmission effect is independent of the detailed features of the defect in that it persists for arbitrary $V$, $\gamma$. 
Comparable transmission resonances at energies where a new scattering channel opens have been observed in transverse-multimode waveguides \cite{Bagwell1990,Baranger1990,Kim1999a}, see also \cite{Saeidian2008,Hess2015}.
The full transmission at $E_c$ is accompanied by $A_1 \rightarrow -1$, so the evanescent waves (whose decay length goes to infinity as $E \rightarrow E_c$) are 
excited at an amplitude identical to that of the incoming travelling wave. Furthermore, $\phi_0 \rightarrow 0$. All these features can be observed in Fig.~\ref{fig:singlek}.\\

\begin{figure}[ht]
\centering
\includegraphics[width=0.42\textwidth]{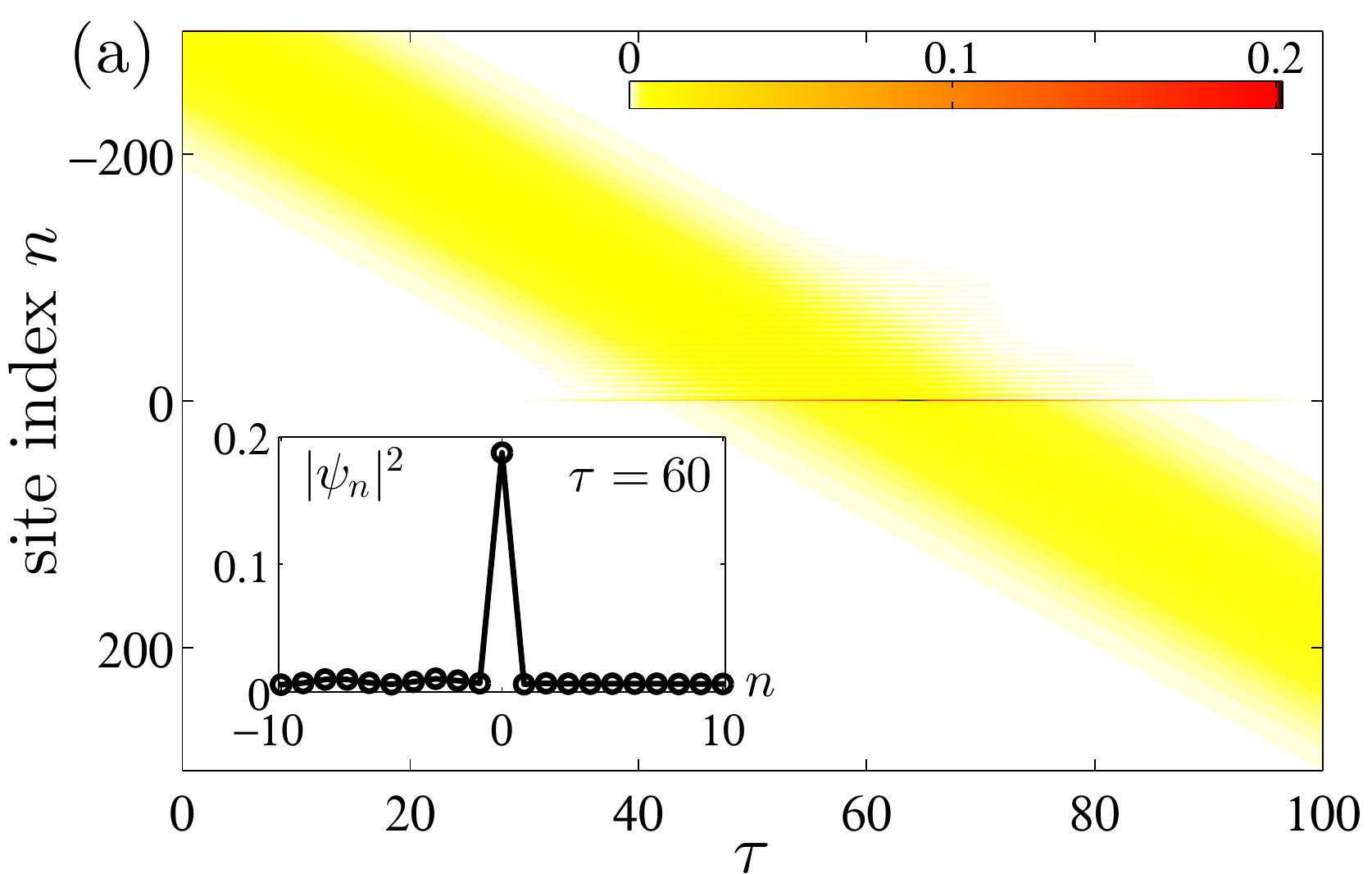} 
\includegraphics[width=0.42\textwidth]{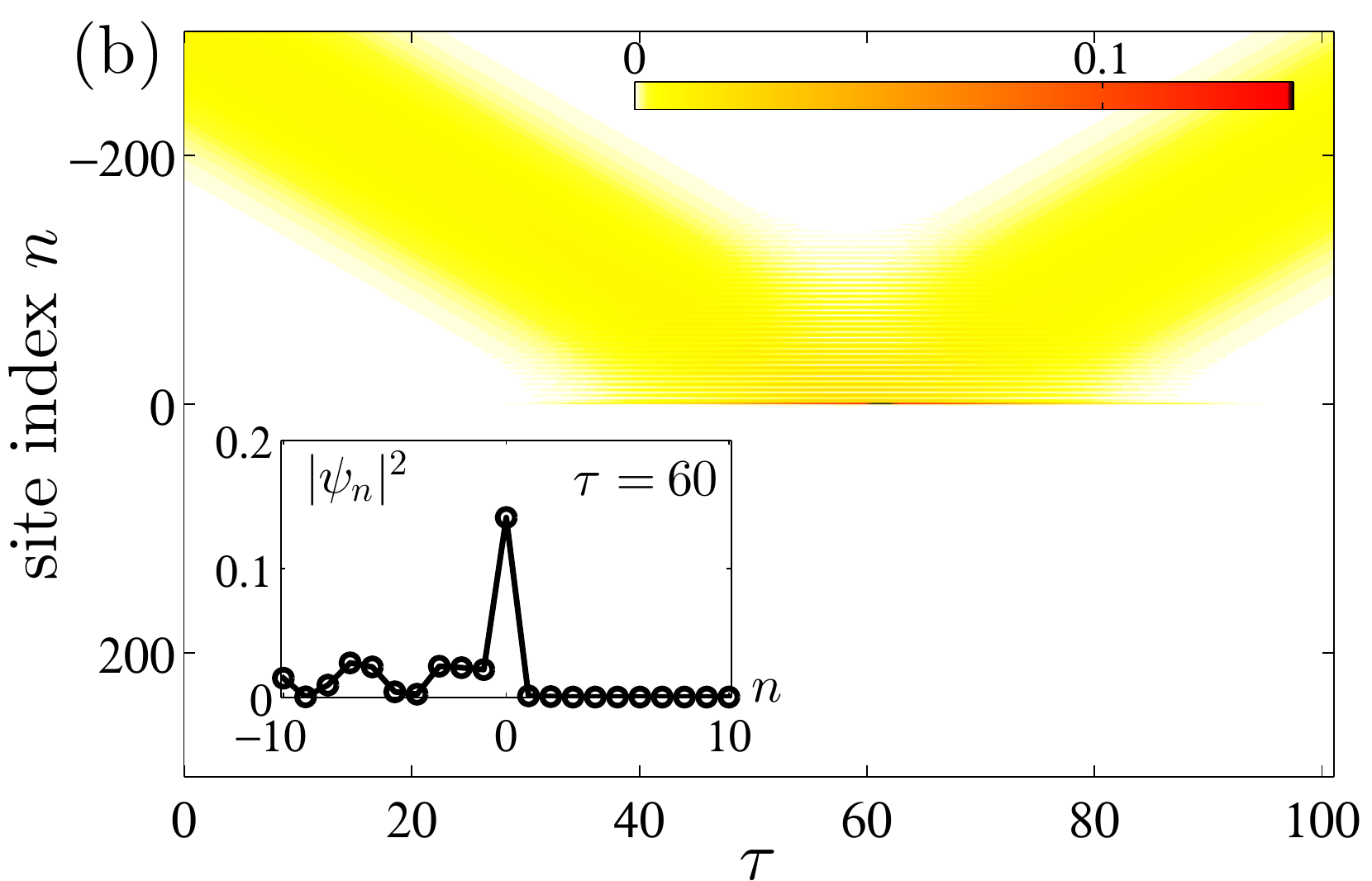} 
\caption{(Color online) Wave packet dynamics at neighboring transmission and reflection resonances for small $\gamma$, the color encodes $|\psi_n|^2$. (a) $k=0.66$, $E =-2.076$. (b) $k=0.71$, $E= -1.817$. Other parameters are $t_1=t_2=1$, $V=-2$, $\gamma=0.2$, $w_0 = 100$, $j_0 = -300$.
The insets show snapshots of the respective densities at time $\tau=60$, demonstrating the resonant excitation of the defect site $n=0$. \label{fig:propfano}}
\end{figure}

Apart from the limiting cases at the edges of the energy interval, it is seen that the transmission probability also becomes unity if $V+E(\gamma^2-1)=0$.
Thus, if the energy $E_T = V/(1-\gamma^2)$ lies in the interval $E_0 < E_T < E_c$, this produces a transmission resonance inside the band. 
Notably, this can only occur for $\gamma \neq 1$. 
Similarly, $p_R =1$ if $V+E(\gamma^2 -1) + \gamma^2 (2t_1 + 8 t_2 \cos k) \sinh \kappa =0$, leading to a reflection resonance inside the band 
if the energy $E_R$ that satisfies this equation lies in the interval $E_0 < E_R < E_c$.\\
In the weak-coupling limit of $\gamma \rightarrow 0$ the latter transmission and reflection resonances both approach $V$.
Thus, if now the defect energy satisfies $E_0 < V < E_c$, for $E \approx V$ the transmission properties of an incoming wave will depend very sensitively on its energy, as is most clearly seen in the corresponding wave packet dynamics.
So far, we have studied stationary scattering solutions. Superimposing these, weighted by a distribution that is localized in $k$-space, 
immediately gives insight into the dynamics of wave packets when scattering from the impurity \cite{Kim2006}.
In particular, for a wave packet that is well localized in $k$-space (and thus broad in direct space), the reflected and transmitted fraction are immediately determined by the 
reflection and transmission probabilities of the stationary scattering solution at the central $k$.
We check this by initializing wide Gaussian wave packets centered at different $k$ near the expected resonance and propagating them towards the defect by direct numerical integration of the discrete Schrödinger equation (\ref{eq:eom}).
Explicitly, the initial condition reads $\psi_j \propto \exp \left[-(j-j_0)^2/w_0^2 + i k j \right]$ with $j_0 \ll 0$ the initial central position and the width parameter $w_0 \gg 1$.
The results are shown in Fig.~\ref{fig:propfano}, demonstrating almost perfect transmission/reflection of the wave packet from the defect upon a small variation of the incoming energy near the resonance,
accompanied by a strong transient excitation of the defect site during the scattering process.
It is in this weak-coupling limit that the zigzag-defect model provides a clean example of the Fano-Feshbach resonance mechanism. We will analyze this in detail in the following.

\begin{figure}[ht]
\centering
\includegraphics[width=0.495\textwidth]{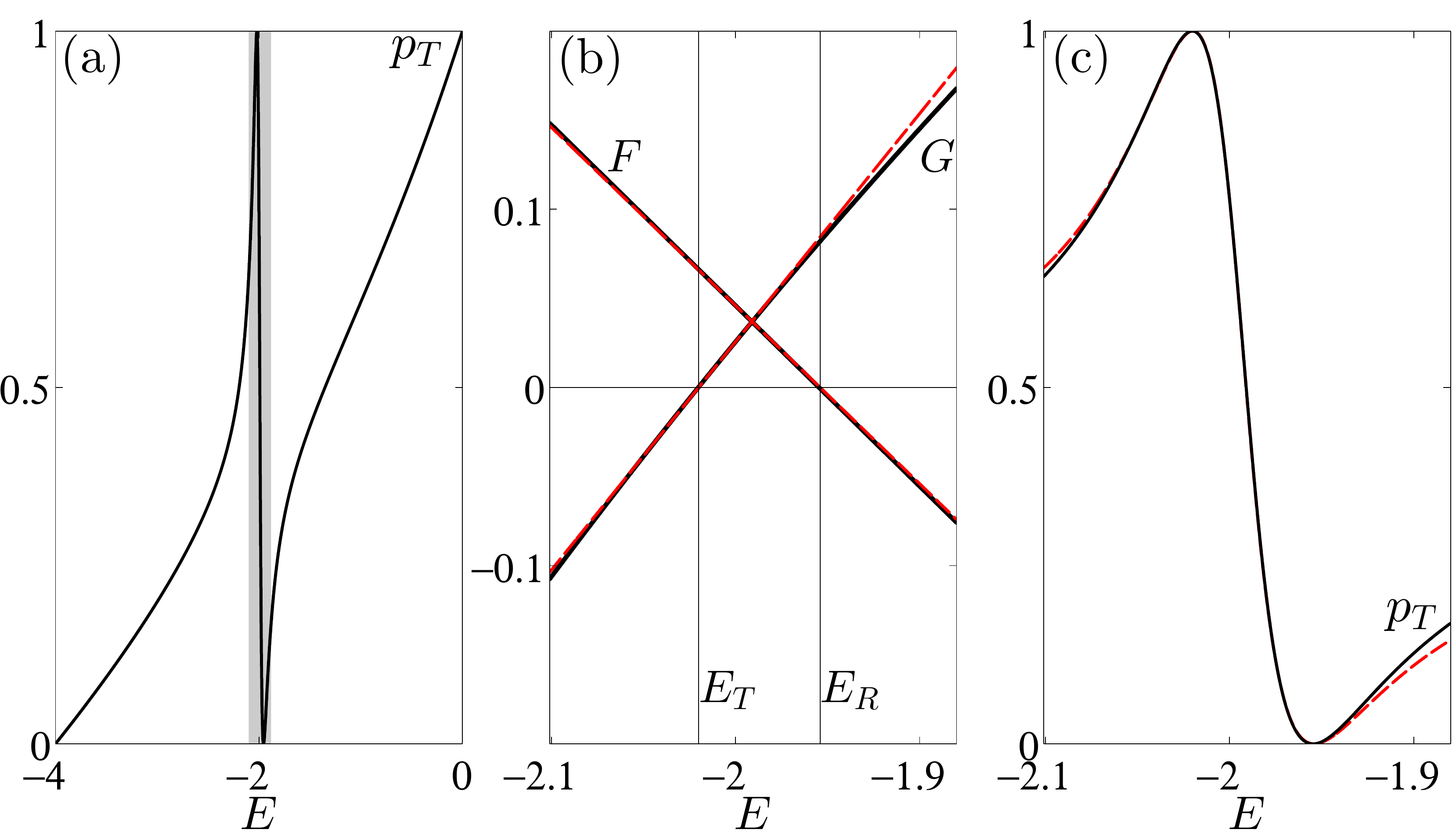} 
\caption{(Color online) Fano resonance at weak coupling,  $t_1=t_2=1$, $\gamma=0.1$, $V=-2$. (a) Transmission probability $p_T$ as a function of the incoming energy in the subcritical interval $E_0 < E < E_c$. 
(b) Functions $F$ and $G$ defined in Eqs.~(\ref{eq:F},\ref{eq:G}) near resonance (solid black lines), together with their zeros $E_R$, $E_T$ and linearizations as predicted from the weak-coupling expansion, Eqs.~(\ref{eq:Fp},\ref{eq:Gp}) (dashed red lines).
(c) Zoom to shaded region of subfigure (a) together with the predicted Fano profile within the weak-coupling expansion, Eqs.~(\ref{eq:fanoprof},\ref{eq:fanoparams}) (dashed red line).\label{fig:fano}}
\end{figure}

\paragraph*{Fano resonance at weak coupling}
To make the connection to the canonical Fano resonance formalism (see e.g. \cite{Joe2006}), it is useful to rewrite the transmission coefficient as $T= F/({F-i G})$,
where
\begin{eqnarray}
 F(E) &=& V+ E(\gamma^2-1) \nonumber \\
 &&+\gamma^2 \left(8 t_2 \cos k + 2 t_1 \right) \sinh \kappa, \label{eq:F} \\
 G(E) &=& -\frac{\sinh \kappa}{\sin k} \left[ V+ E(\gamma^2 -1) \right] \label{eq:G}
\end{eqnarray}
are real functions, cf. \cite{Tribelsky2008}. Then, the aforementioned transmission/reflection resonances $E_T, E_R$ occuring near $V$ at weak coupling correspond to the zeros of $G$ and $F$, respectively:
$G(E_T)=0$, $F(E_R)=0$.
To lowest non-vanishing order in the coupling parameter $\gamma^2$, one can expand the zeros as
\begin{eqnarray*}
 E_T &\approx& V + \gamma^2 V, \\
 E_R &\approx& V + \gamma^2 V \\
&+& \gamma^2 8 t_2\left( \cos k_V + \frac{t_1}{4 t_2} \right) \sqrt{ \left(\frac{t_1}{2 t_2} + \cos k_V \right)^2 -1} ,
\end{eqnarray*}
where
\begin{equation*}
 \cos k_V = -\frac{t_1}{4 t_2} + \sqrt{ \left( \frac{t_1}{4t_2} \right)^2- \frac{V}{4 t_2} + \frac{1}{2} }.
\end{equation*}
In a second step, one can linearize $F$ and $G$ around their respective zeros, assuming that these zeros lie close enough to each other
such that there is a common region where both linearizations apply (which is certainly true in the limit of $\gamma^2 \rightarrow 0$ in which both zeros coincide and will approximately still hold for small enough $\gamma^2$).
Then the transmission coefficient in the vicinity of the resonances is given by
\begin{eqnarray*}
 T = \frac{  F'(E_R) (E-E_R)} { F'(E_R)(E-E_R) -i G'(E_T) (E-E_T)}
\end{eqnarray*}
within the weak-coupling expansion, where the prime denotes a derivative with respect to $E$. This yields the well-known Fano profile
\begin{equation}
p_T(E) =|T(E)|^2 = \frac{1}{1+q^2} \frac{(E-\bar{E} +q \Gamma/2)^2}{(\Gamma/2)^2 +(E-\bar{E})^2}
\label{eq:fanoprof}
\end{equation}
upon the identifications
\begin{equation}
 q=\frac{G'(E_T)}{F'(E_R)}, \, \bar{E} = \frac{E_R + q^2 E_T }{1+q^2}, \, \Gamma = 2q \frac{E_T-E_R}{1+q^2}.
\label{eq:fanoparams}
\end{equation}
To evaluate this, we calculate the zeroth order contributions of the derivatives which we find to be
\begin{eqnarray}
 F'(E_R) &\approx& -1, \label{eq:Fp}\\
 G'(E_T) &\approx& \sqrt{ \frac{ t_1^2 + 4 t_1 t_2 \cos k_V}{4t_2^2 \left(1- \cos^2 k_V \right)} -1} \label{eq:Gp}.
\end{eqnarray}
To lowest order in the coupling, this gives for the Fano asymmetry parameter
\begin{equation}
 q = - \left[ \frac{ t_1 \sqrt{ t_1^2-4 t_2 V + 8 t_2^2} - 4t_2^2-2 V t_2 + t_1^2  }{t_1 \sqrt{ t_1^2-4 t_2 V + 8 t_2^2} + 4t_2^2 +2 V t_2 - t_1^2  }\right]^{1/2}.
\label{eq:fanoq}
\end{equation}
Corresponding approximate values for the resonance position $\bar E$ and width $\Gamma$ can then be obtained by inserting this into the expressions of Eq.~(\ref{eq:fanoparams}).
The ensuing prediction of the Fano resonance profile, Eq~(\ref{eq:fanoprof}), is compared to the exact transmission line in Fig.~\ref{fig:fano} for an example set of parameters, showing very good agreement.
As expected, in the weak-coupling limit the width $\Gamma$ is of order $\gamma^2$, thus producing the characteristic narrow resonance structure.

\begin{figure}[ht]
\centering
\includegraphics[height=0.2\textwidth]{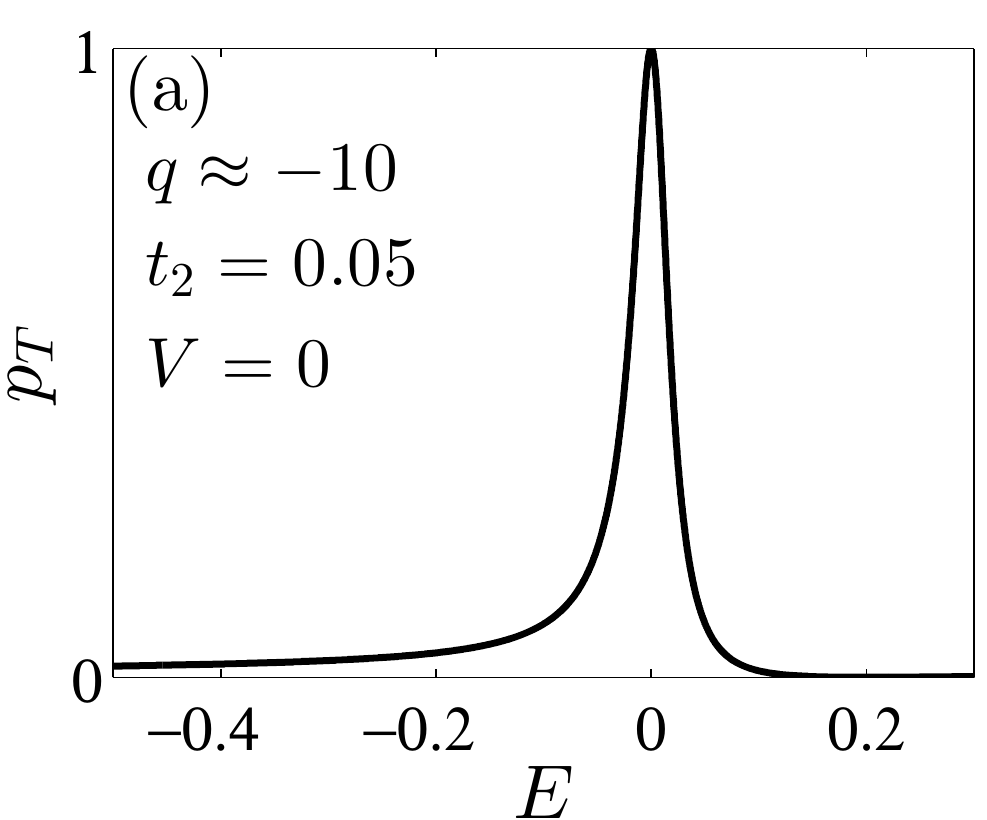} 
\includegraphics[height=0.2\textwidth]{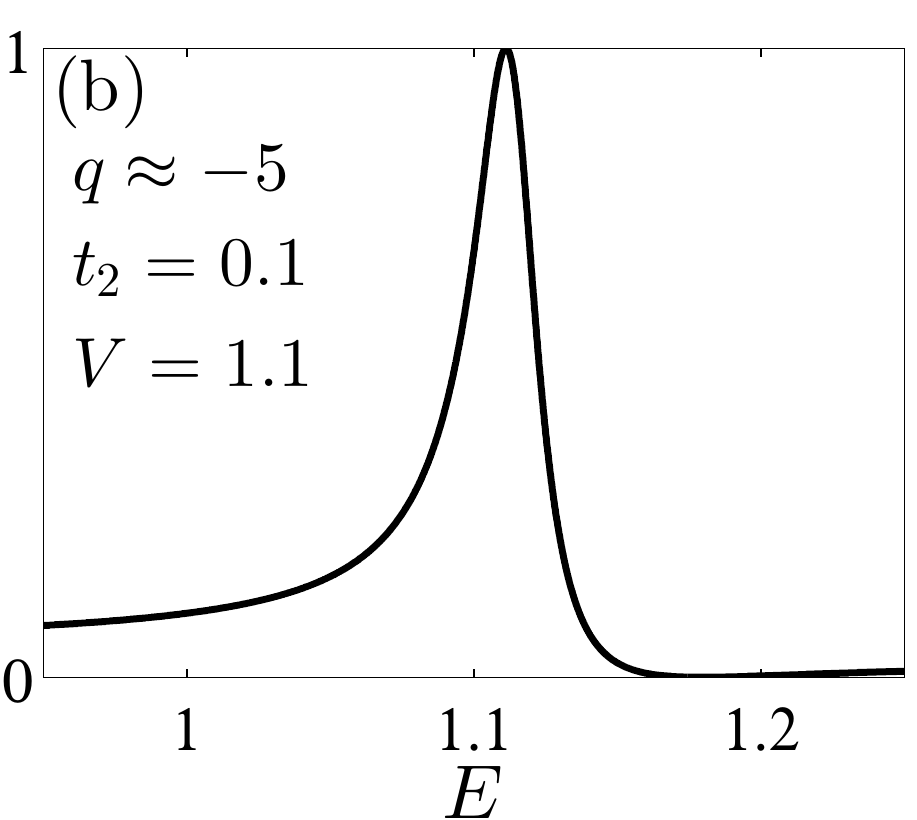} \\
\includegraphics[height=0.2\textwidth]{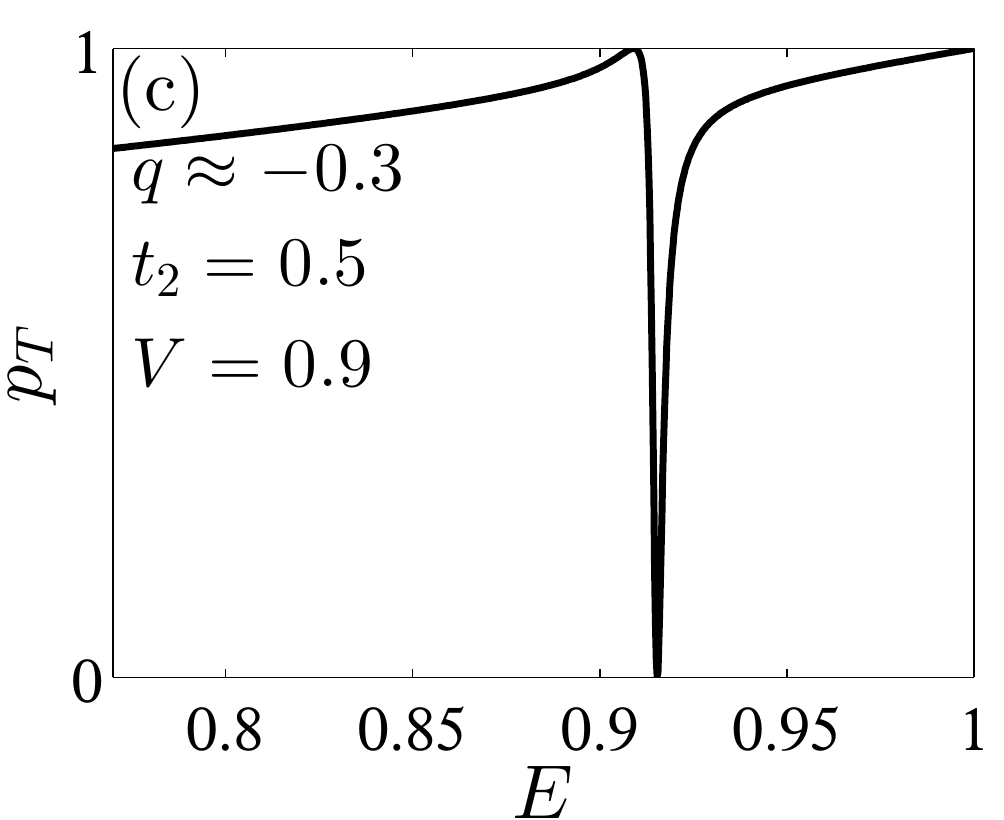} 
\includegraphics[height=0.2\textwidth]{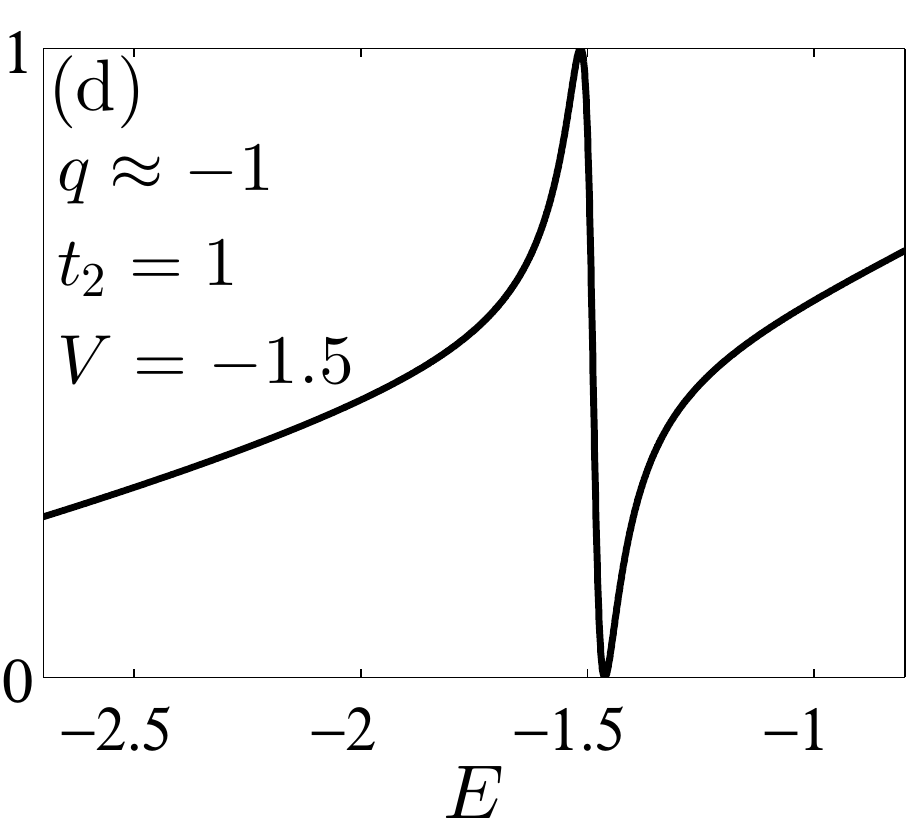} 
\caption{Transmission probability $p_T$ as a function of energy $E_0 < E < E_c$ in the weak-coupling regime, demonstrating tunability of the Fano $q$-parameter via $t_2, V$. Throughout $t_1 = 1$, $\gamma=0.1$, $t_2$ and $V$ as given in the figures.  
The indicated values of $q$ are calculated from Eq.~(\ref{eq:fanoq}) and agree with the result from a Fano profile fit to the respective resonance region on the percent level.
\label{fig:fanoq}}
\end{figure}

Eq.~(\ref{eq:fanoq}) suggests a high degree of tunability of the Fano lineshape through the asymptotic hopping parameters $t_1, t_2$ of the lattice and the defect potential $V$.
First, we note that in this system $q < 0$ throughout, thus there is no $q$-reversal even when $V$ crosses zero.
The absolute value of $q$, on the other hand, is subject to large variations as illustrated in Fig.~\ref{fig:fanoq}.
When concentrating on small $t_2$ and $|V|$, we find the limiting case discussed in \cite{Tribelsky2008} in which $q \ll -1$ and the Fano profile approaches a symmetric Breit-Wigner lineshape [Fig.~\ref{fig:fanoq}(a)].
The same is found more generally when $V \gtrsim E_0$ and thus the resonance occurs at small $k$.
Interestingly, in the opposite limit of $V \lesssim E_c$ the limiting value of $q$ crucially depends on the ratio of the hopping parameters:
If $t_2<t_1/4$, then again $q \ll -1$, eventually approaching a symmetric Breit-Wigner lineshape [Fig.~\ref{fig:fanoq}(b)].
But if $t_2 > t_1/4$, we find that $q \rightarrow 0$ instead, asymptoting towards an inverted Breit-Wigner shape [Fig.~\ref{fig:fanoq}(c)]. 
This qualitative change in $q$ reflects the qualitative change of the transmission probability at $E_c$ for the two cases 
(full transmission for $t_2 > t_1/4$, zero transmission for $t_2 < t_1/4$).
For the former case, the resonance reduces to an approximately symmetric dip on top of a background of transmission unity, while for the latter case it corresponds to an approximately symmetric spike of full transmission on a zero background.
In contrast, a maximally asymmetric Fano lineshape with $q=-1$ is obtained when $V=t_1^2/(2 t_2) -2t_2$ [Fig.~\ref{fig:fanoq}(d)].

\section{Scattering at supercritical energies}
\label{sec:super}
\begin{figure}[ht]
\includegraphics[width=0.47\textwidth]{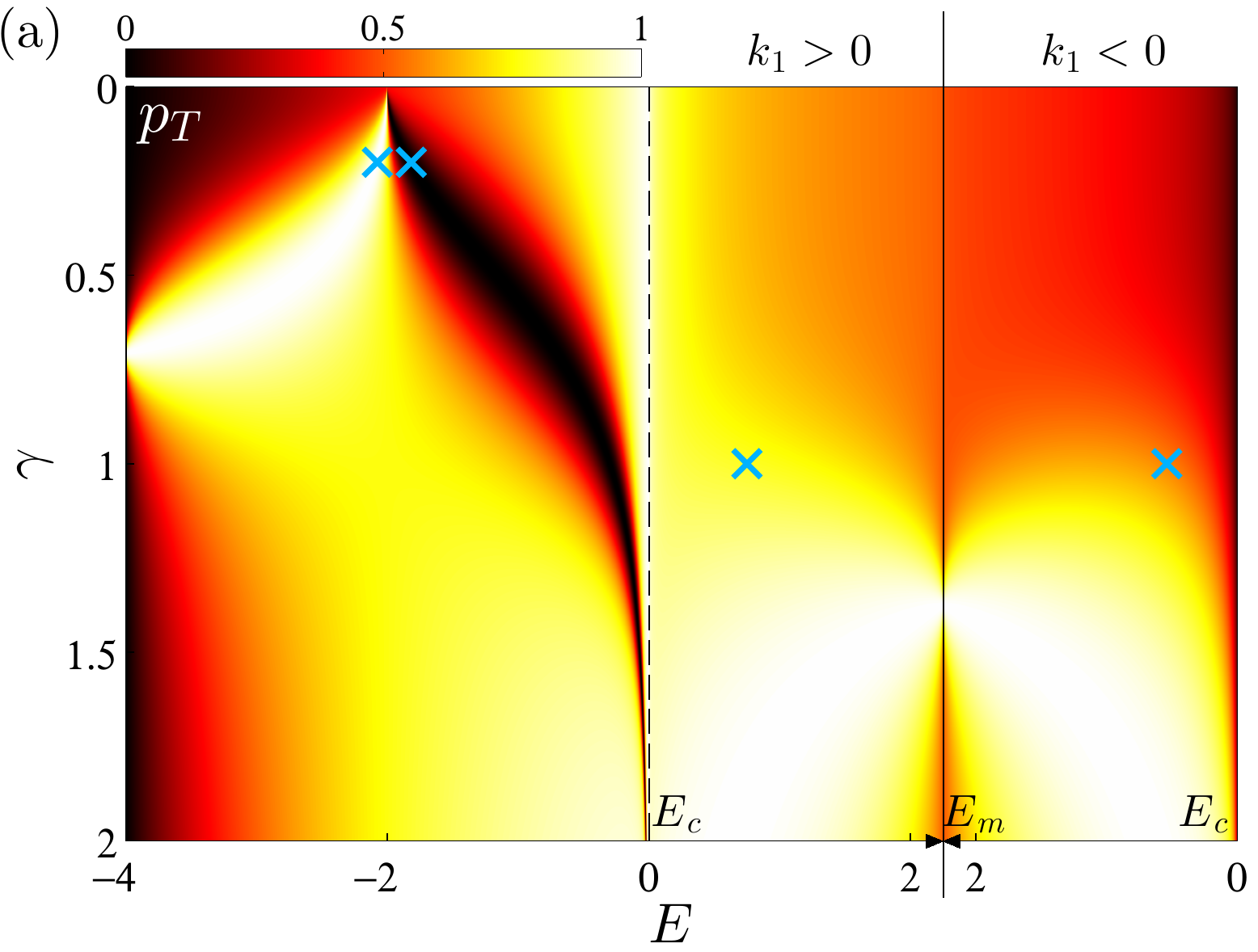}\\
\includegraphics[width=0.235\textwidth]{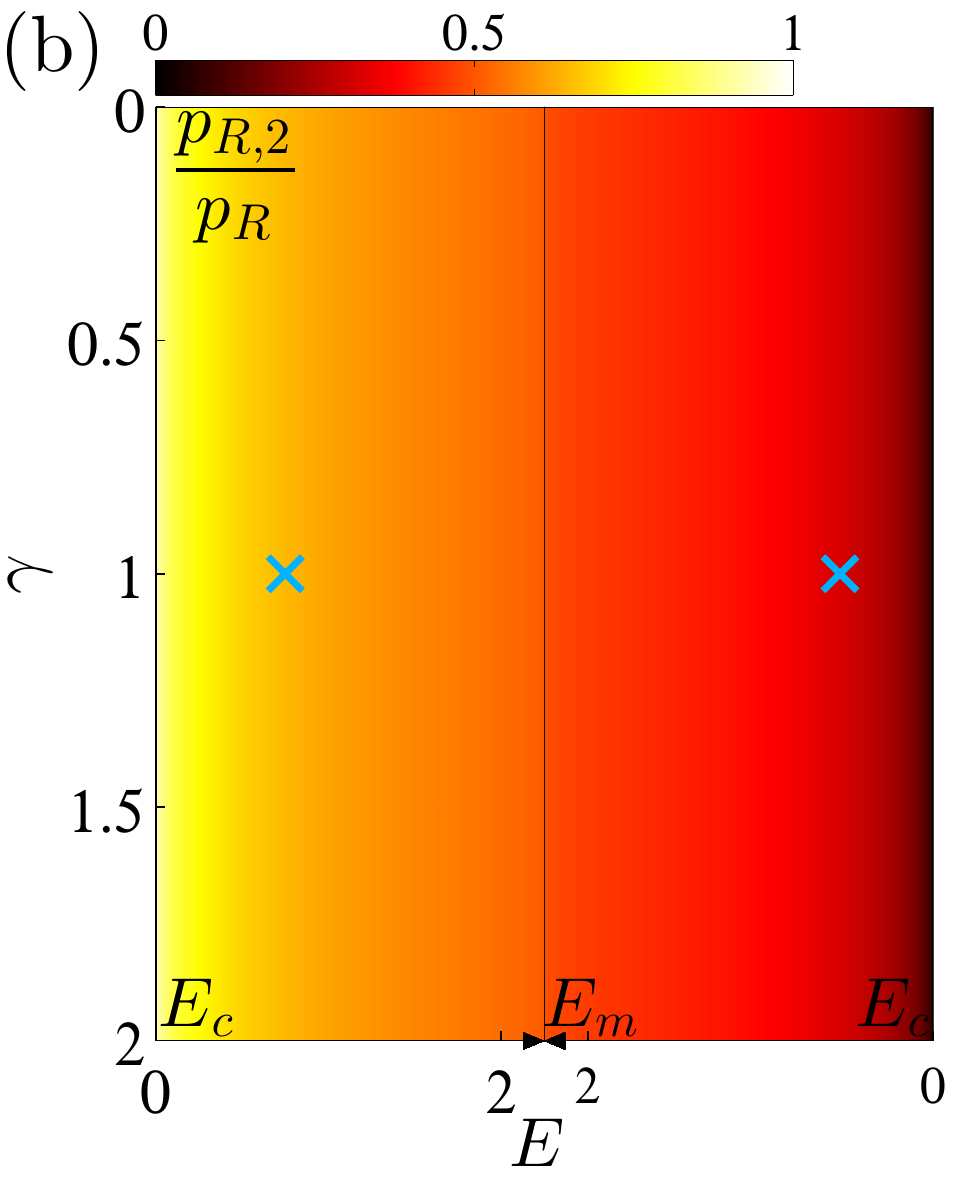}
\includegraphics[width=0.235\textwidth]{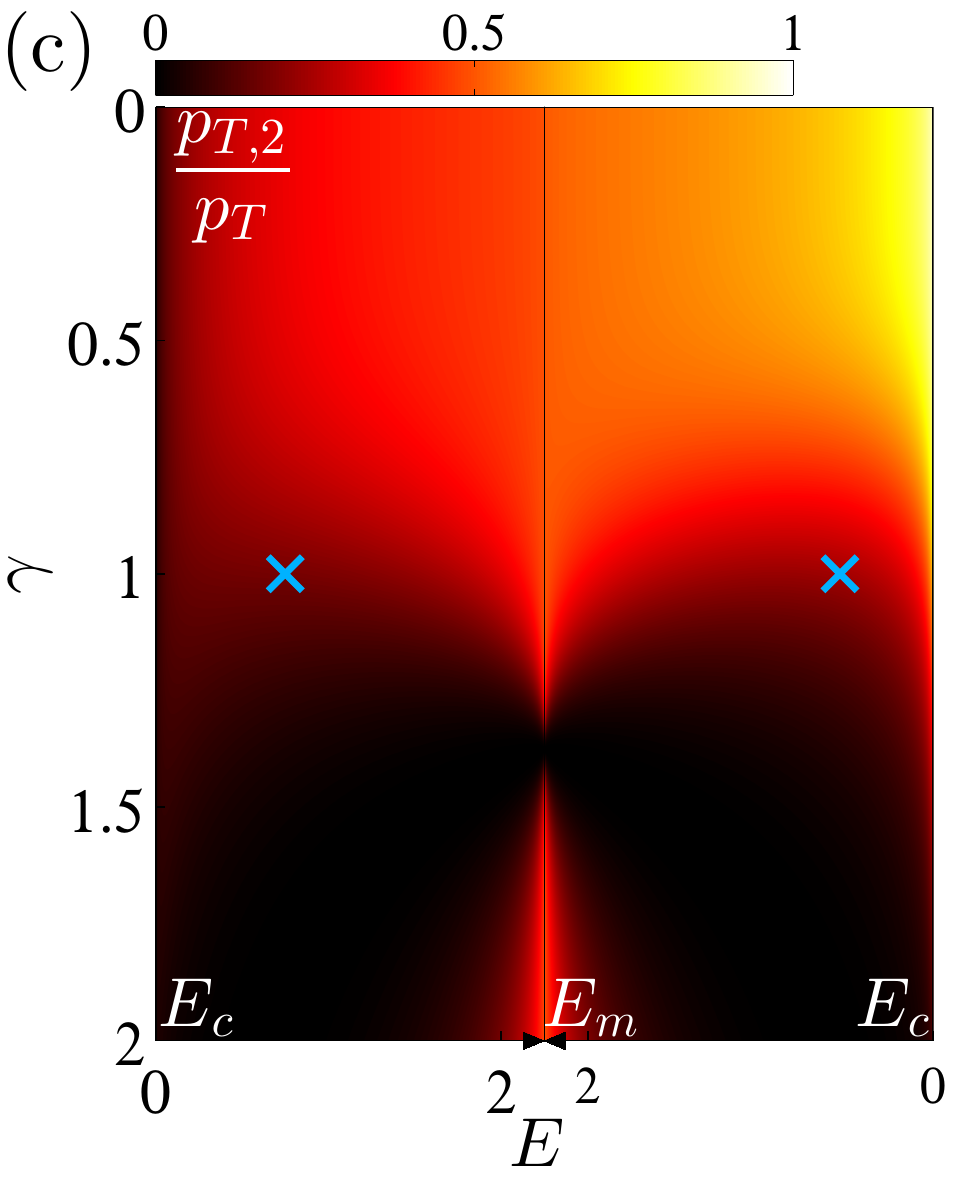}
\caption{(Color online) (a) Total transmission as a function of the incoming energy $E$ and the coupling parameter $\gamma$. 
The degenerate regions in the supercritical regime are distinguished by the sign of $k_1$, see discussion in the text. Note that above the critical energy $E_c$ the energy axis is nonmonotonic, increasing towards $E_m$ and then decreasing to $E_c$ again.
(b,c) Branching ratios of reflection/transmission into the second channel to the total reflection/transmission. The vertical lines at $E=E_m$ separate the regions of $k_1 > 0$ and $k_1 <0$, as in (a).
Throughout, $t_1=t_2=1, V=-2$. Crosses mark the parameter values at which wave packet propagation runs are shown in Figs.~\ref{fig:propfano} and \ref{fig:prop2}.\label{fig:trans1}}
\end{figure}

We now consider scattering from the defect for energies at which two different propagating waves coexist.
As discussed before, this can only happen if $t_2 > t_1/4$ and $E_c < E < E_m$, which is assumed throughout this section.
Let $k_1$ denote the wave number of the incoming wave and $k_2$ (of the same sign and thus opposite group velocity) the second solution with $E(k_1)=E(k_2)=E$.
The scattering Ansatz is now
\begin{eqnarray*}
  \phi_j &=& e^{i k_1 j} + R_1 e^{-i k_1 j} + R_2 e^{i k_2 j}, \qquad j < 0,\\
  \phi_j &=& T_1 e^{i k_1 j} +T_2 e^{-i k_2 j}, \qquad \qquad \quad \,\,\, j >0,
\end{eqnarray*}
where again $\phi_0$ is an additional unknown. The terms with prefactors $R_i$, $T_i$ correspond to reflected/transmitted waves in the two channels, respectively, 
while only the $\exp(i k_1 j)$ wave is incoming.
Note the different choices of signs accordings to the different signs of the group velocities. 
This Ansatz assumes $v(k_1)>0$ and $v(k_2)<0$, thus one of the following two cases must hold, see Fig.~\ref{fig:BSDOS}.
Either $k_1 \in (k_c,k_m)$ and  $k_2 = \text{acos}\left[-t_1/(2t_2)-\cos k_1 \right] \in (k_m,\pi)$,
or $k_1 \in (-\pi,-k_m)$ and $k_2 = -\text{acos}\left[-t_1/(2t_2)-\cos k_1 \right] \in (-k_m,-k_c)$.
\\
Inserting the scattering Ansatz into Eqs.~(\ref{eq:eomset}) and solving the resulting inhomogeneous linear equation now yields
\begin{eqnarray}
 R_1&=& \left[ i \frac{\gamma^2 v(k_1)}{V+E(\gamma^2-1)} - \frac{\sin k_1 + \sin k_2}{\sin k_2} \right]^{-1} \label{eq:R1},\\
 R_2&=& T_2 = \frac{\sin k_1}{\sin k_2} R_1, \quad T_1 = R_1 + 1, \label{eq:R2}\\
 \phi_0 &=& (R_1 + R_2 +1)/\gamma \label{eq:phi02}.
\end{eqnarray}
Evaluating the continuity equation (\ref{eq:continuity2}), we find the following relation reflecting probability conservation
\begin{equation}
 \underbrace{|T_1|^2}_{p_{T,1}} + \underbrace{|R_1|^2}_{p_{R,1}}+  \underbrace{\left| \frac{v(k_2)}{v(k_1)} \right| |T_2|^2}_{p_{T,2}} + \underbrace{\left| \frac{v(k_2)}{v(k_1)} \right| |R_2|^2}_{p_{R,2}} =1,
\label{eq:contsup}
\end{equation}
which can be checked to be satisfied by the above explicit expressions. It is worth noting that for the two different wave numbers of the same energy the ratio of the group velocities is given by the simple formula
\begin{equation}
 \frac{v(k_2)}{v(k_1)} = - \frac{\sin k_2}{\sin k_1}, \quad \text{if } E(k_1)=E(k_2), \, |k_1| \neq |k_2|.
\end{equation}
Formally, replacing $k_2 \rightarrow -\pi -i \kappa$, Eqs.~(\ref{eq:R})--(\ref{eq:A1}) can be recovered from Eqs.~(\ref{eq:R1})--(\ref{eq:phi02}).\\
Fig.~\ref{fig:trans1} shows the total transmission probability $p_T=p_{T,1}+p_{T,2}=1-p_R$ as a function of the incoming energy $E$ when varying the coupling $\gamma$, while keeping $t_1$, $t_2$, $V$ fixed.
We include here both subcritial energies $E_0 < E < E_c$ (where there is only one open channel) and supercritical energies $E_c < E < E_m$. In the latter range, specifying the energy $E$ does not uniquely fix the incoming wave number 
(as per our above discussion of the degeneracy of the band structure curve). The two different initial quasi-momenta corresponding to the same $E$ here can be distinguished by their sign, $k_1 > 0$ or $k_1 < 0$.
First, it can again be seen here that at $E_c$ the transmission probability is unity.
In the weak-coupling regime, we clearly observe the Fano resonance structure at an energy near $V$, as discussed before.
Increasing the coupling, the transmission and reflection resonances are shifted away from $V$. The reflection resonance $E_R$ moves to larger $E$. 
Depending on the details of the parameters, it may reach $E_c$ at a finite $\gamma^2$ and then cease to exist (this happens
at $\gamma^2=1-V/E_c$ if the latter quantity is positive). Interestingly, in the case shown here, with $E_c = 0$, the reflection resonance does not vanish at a finite $\gamma^2$, 
but instead becomes very narrow and asymptotes to $E_c$ (where, as noted before, full transmission is found).
Thus, also at stronger couplings we identify a region in which the transmission probability sensitively depends on the incoming energy (similar to the Fano resonance interval at weak coupling), namely for energies just below $E_c$. 
The fate of the transmission resonance $E_T = V/(1-\gamma^2)$ is mainly determined by the sign of $V$. For $V<0$, as is the case in Fig.~\ref{fig:trans1}, it drifts towards the lower band edge $E_0$ and disappears there at $\gamma^2 = 1-V/E_0$.
In cases with $1-V/E_c$ finite and positive, the transmission resonance reappears at $E_c$ for larger $\gamma^2$. This does not happen here due to $E_c = 0$ in the example. For $0<V<E_c$, $E_T$ first drifts towards $E_c$ with increasing $\gamma^2$ instead.\\

\begin{figure}[ht]
\centering
\includegraphics[width=0.45\textwidth]{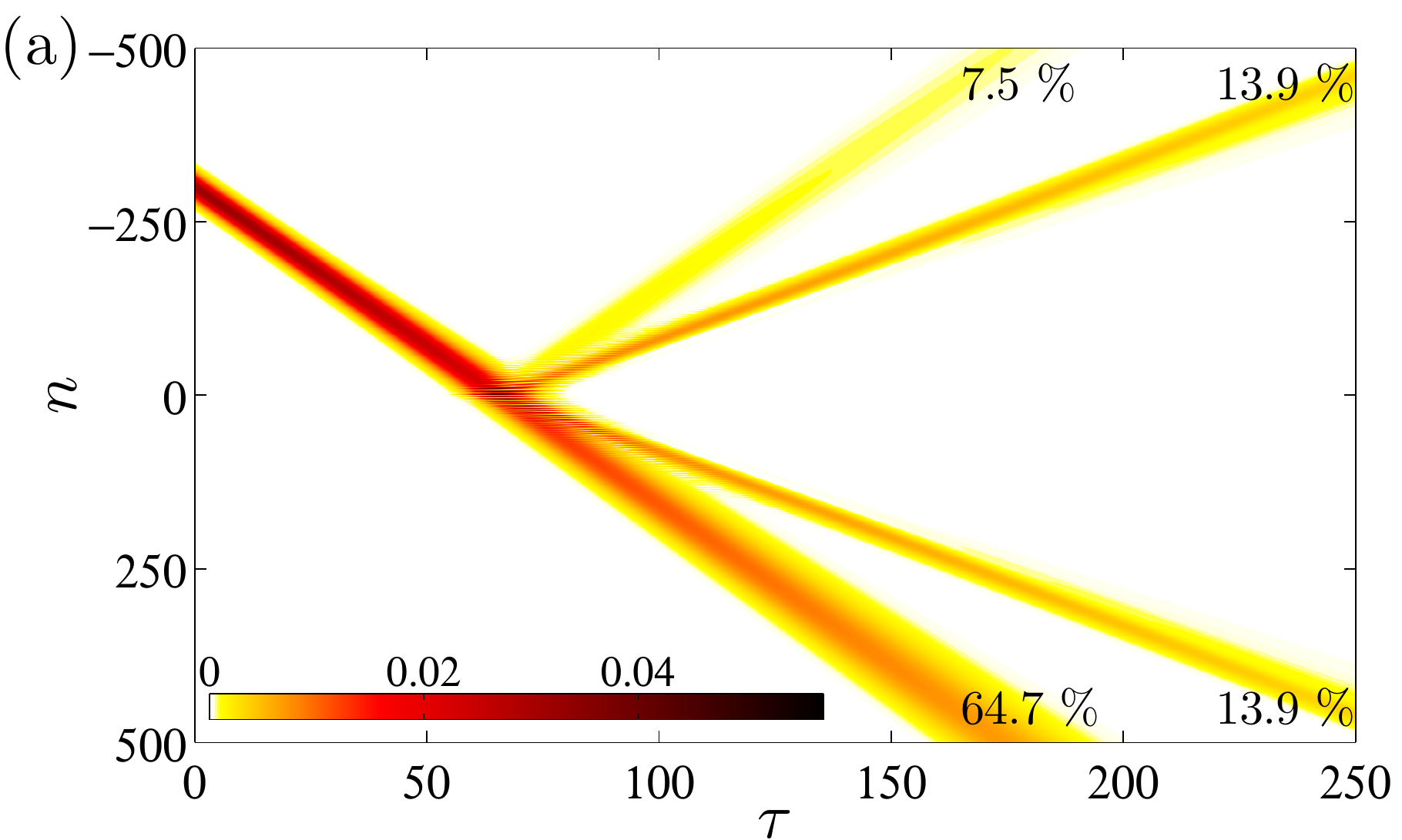} 
\includegraphics[width=0.45\textwidth]{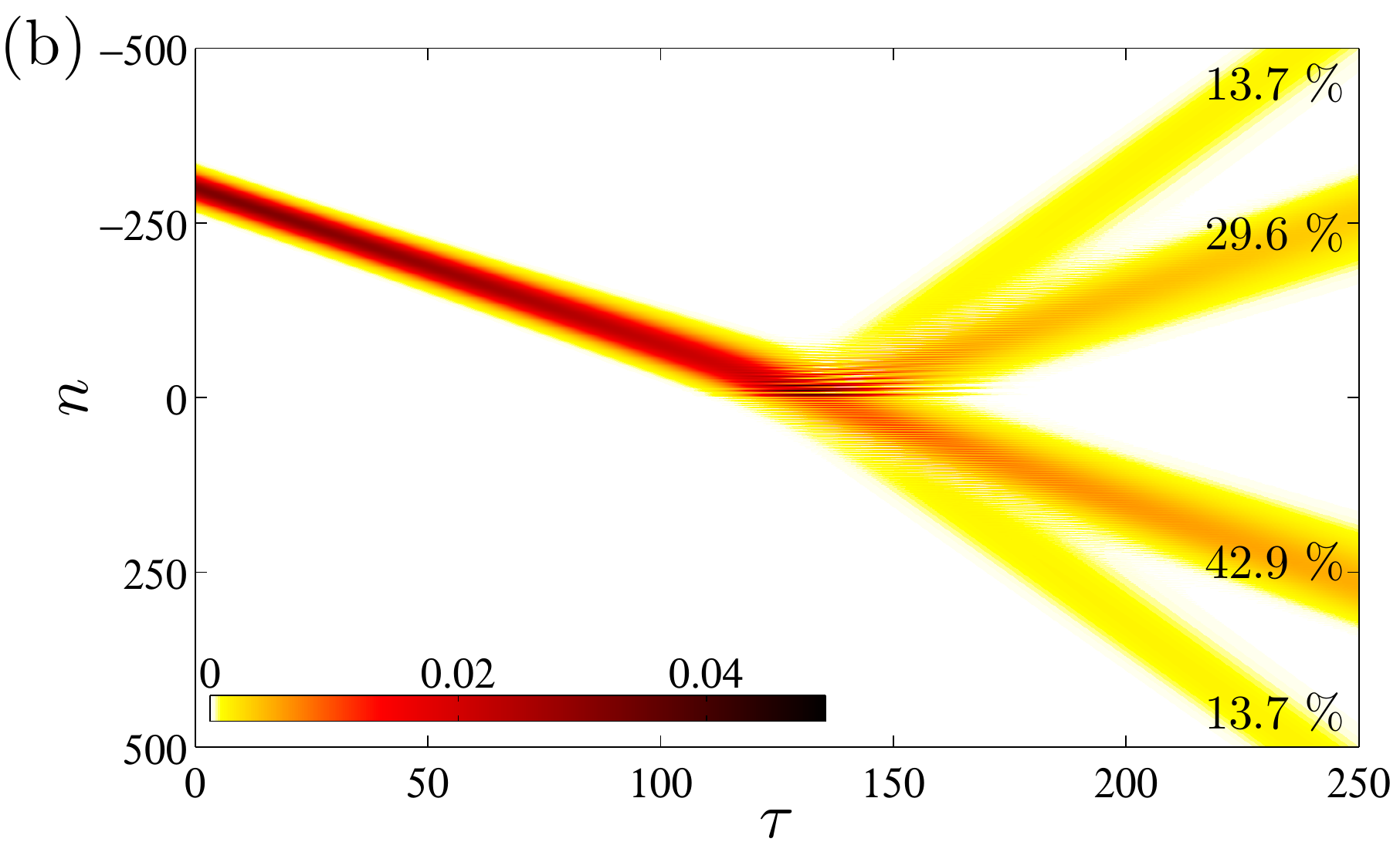} 
\caption{(Color online) Wave packet dynamics in the supercritical regime, the color encodes $|\psi_n|^2$. (a) $k_1=1.2$, $E =0.75$. (b) $k_1=-2.7$, $E= 0.539$. Other parameters: $t_1=t_2=1$, $\gamma=1$, $V=-2$, $j_0 = -300$, $w_0 = 25$. \label{fig:prop2}}
\end{figure}

\begin{figure}[ht]
\centering
\includegraphics[width=0.45\textwidth]{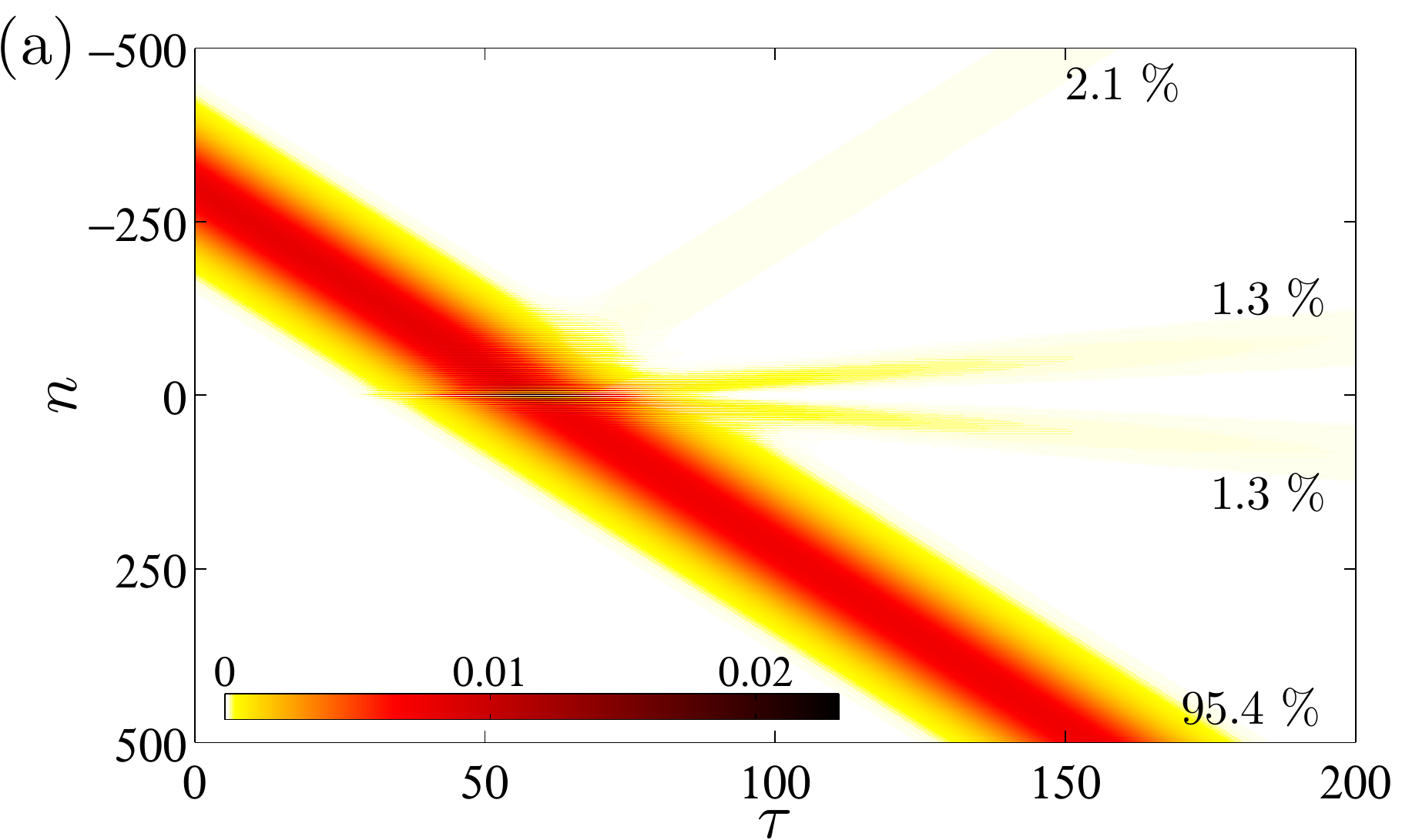} 
\includegraphics[width=0.45\textwidth]{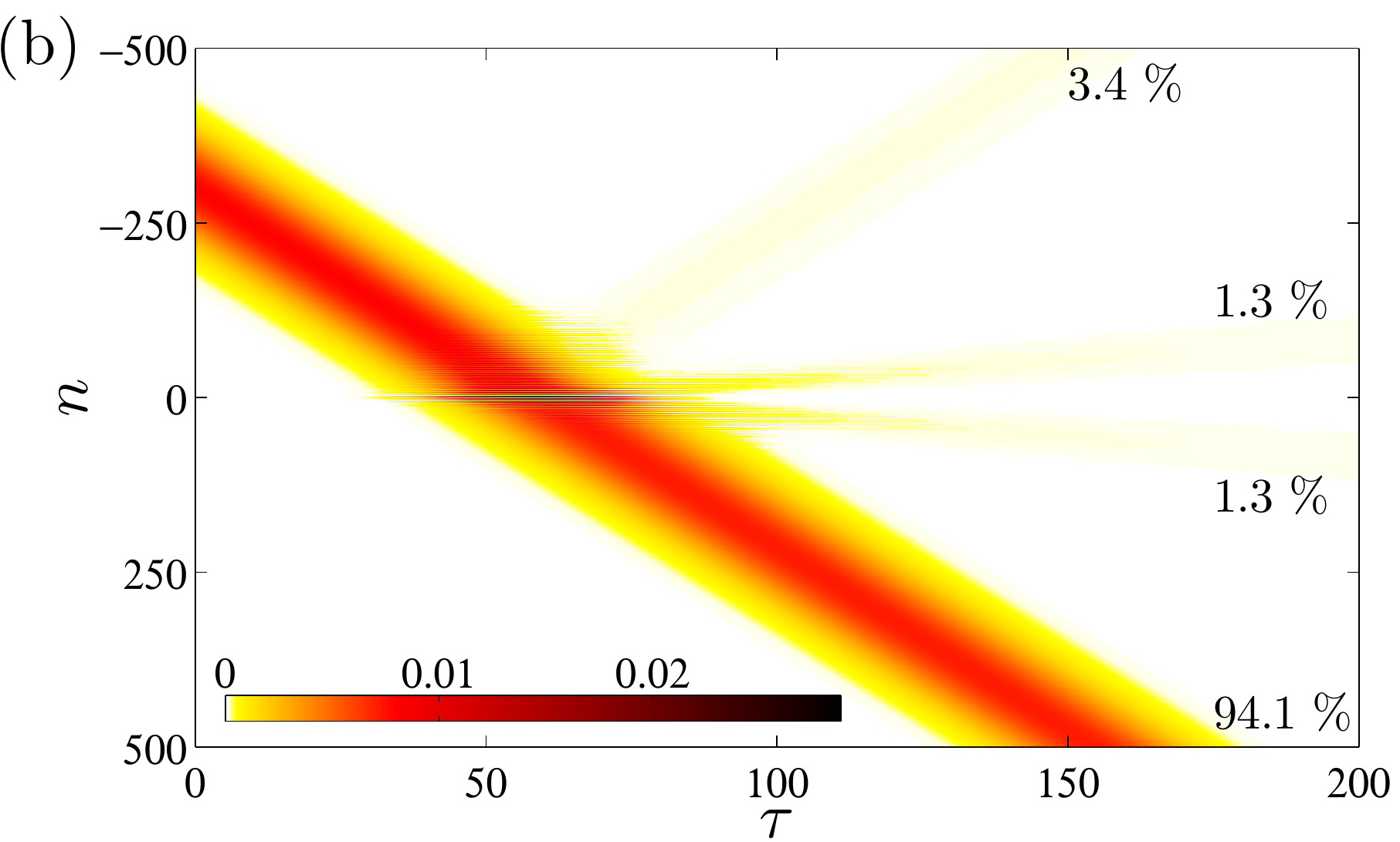} 
\caption{(Color online) Almost full wave packet transmission when approaching the critical energy, $k=k_c-0.01$. The color encodes $|\psi_n|^2$. (a) $V=+5$. (b) $V=-5$. Other parameters are: $t_1=t_2=1$, $\gamma=0.8$, $j_0 = -300$, $w_0 = 100$. \label{fig:prop3}}
\end{figure}

In addition to the total transmission probability, Fig.~\ref{fig:trans1} also displays the respective contributions of the second channel in the reflection and the transmission, respectively, see Eq.~(\ref{eq:contsup}).
These can be simplified to
\begin{eqnarray}
 &&\frac{p_{R,2}}{p_R} = \left( 1 + \frac{p_{R,1}}{p_{R,2}} \right)^{-1} = \left( 1 + \left| \frac{ \sin k_2}{ \sin k_1} \right| \right)^{-1}, \\
 &&\frac{p_{T,2}}{p_T} = \frac{1-p_T}{p_T}  \left( 1 + \left| \frac{ \sin k_2}{ \sin k_1} \right| \right)^{-1}.
\end{eqnarray}
Remarkably, the contribution of the second channel to the reflected fraction is thus independent of the defect parameters $V$, $\gamma$ and only depends on the incoming momentum $k_1$ and the asymptotic parameters $t_1$, $t_2$ of the lattice (which determine the corresponding $k_2$). In Fig.~\ref{fig:trans1}, it is clearly seen that $p_{R_2}/p_R$ is independent of $\gamma$ (see also Fig.~\ref{fig:trans2} below). In contrast, the contribution of the second channel to the transmitted wave does depend on the overall transmission, and thus on the defect details.\\
There are regions of parameter space in which ${p_{R,2}}/{p_R}$ or ${p_{T,2}}/{p_T}$ approach 1, i.e. where the reflected/transmitted wave is dominated by the second channel (e.g., most prominently for the reflected part at $k_1 \gtrsim k_c$). 
This, however, tends to be accompanied by small overall reflection/transmission: Only in regions where just a small part of the incoming wave is reflected/transmitted will the reflected/transmitted part have dominant contributions from the second channel.
For more generic parameter values with incoming energies $E_c < E < E_m$, there are nonvanishing contributions of both the first (incoming) and the second channel in both the transmitted and the reflected part. 
This is observed in the wave packet propagation runs shown in Fig.~\ref{fig:prop2}.
A wave packet of central momentum $k_1$ with $E(k_1) > E_c$ is initialized in the asymptotic region of the lattice. It propagates towards the defect at the group velocity $v(k_1)$. After scattering from the defect, there are four wave packets, two corresponding to the incoming channel and propagating at $v(k_1)$ (transmitted) and $-v(k_1)$ (reflected), respectively.
In addition, two new wave packets are emitted symmetrically from the collision event, equal in shape (according to $T_2=R_2$) and travelling at the velocities $\pm v(k_2)$.
Fig.~\ref{fig:prop2} also displays the final contributions of the four individual wave packets to the total norm. These agree with the weights in Fourier space, as expected.
When the initial wave packet is wide enough to be considered essentially localized in momentum space, the weights of the individual wave packets after the separation agree well with the reflection/transmission probabilities extracted from the stationary scattering solutions (within below 1\% for the runs shown in the figure). We find that in all cases considered $p_{R,2}+p_{T,2} \leq 0.5$, thus the transfer to the second channel does not exceed 50\%.
A qualitative difference between the regimes with incoming $k_1 > 0$ and $k_1 < 0$ is worth noting. In the former case, we have $|v(k_1)| > |v(k_2)|$, so the secondary wave packets created in the scattering process lag behind those travelling at the velocity of the incoming wave packet [Fig.~\ref{fig:prop2}(a)]. In the latter case of $-\pi < k_1 < -k_m$, we find $|v(k_2)| > |v(k_1)|$ instead, so the emerging wave packets at momenta $\pm k_2$ travel at a velocity faster than the initial one [Fig.~\ref{fig:prop2}(b)].\\
\begin{figure}[ht]
\includegraphics[width=0.47\textwidth]{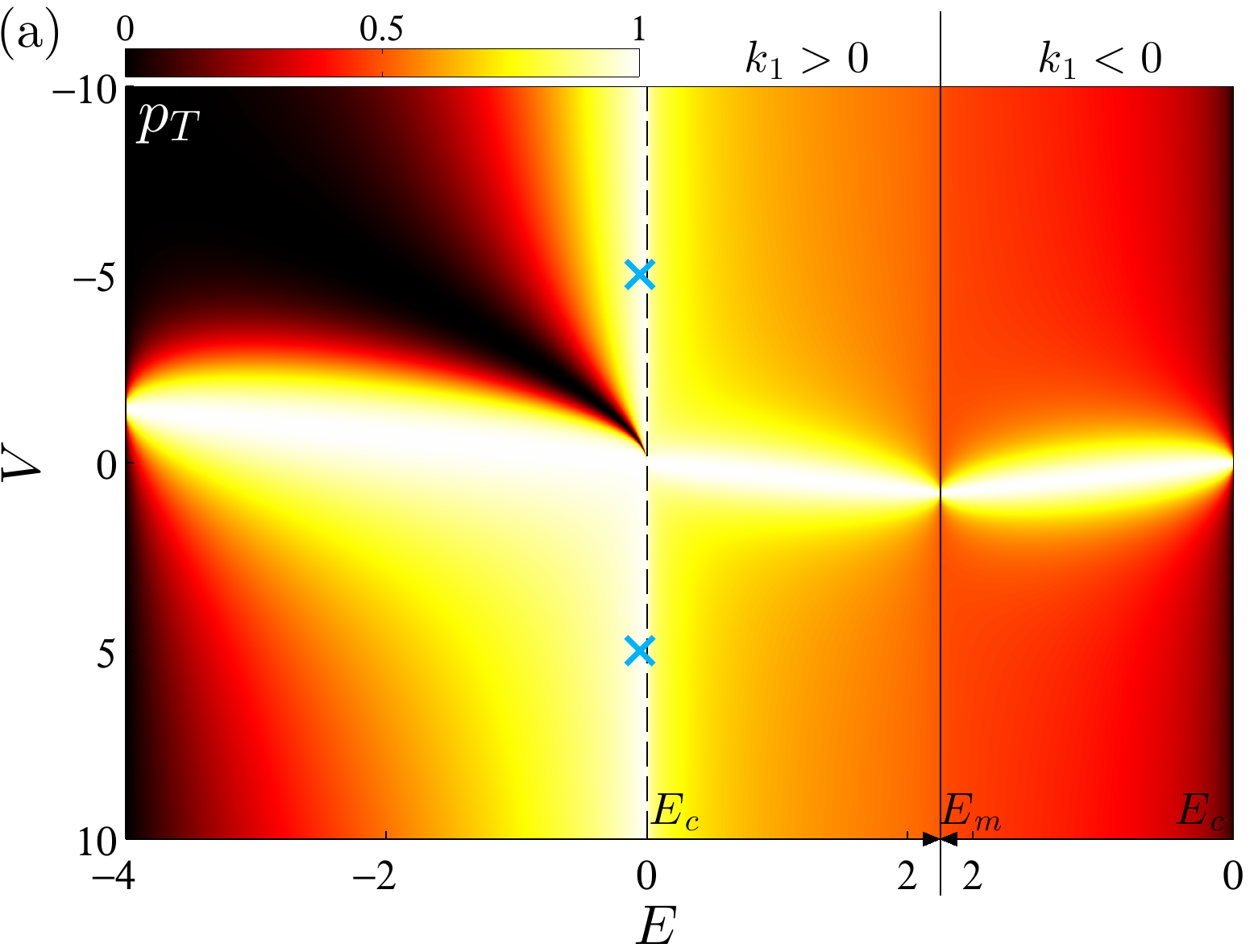}\\
\includegraphics[width=0.235\textwidth]{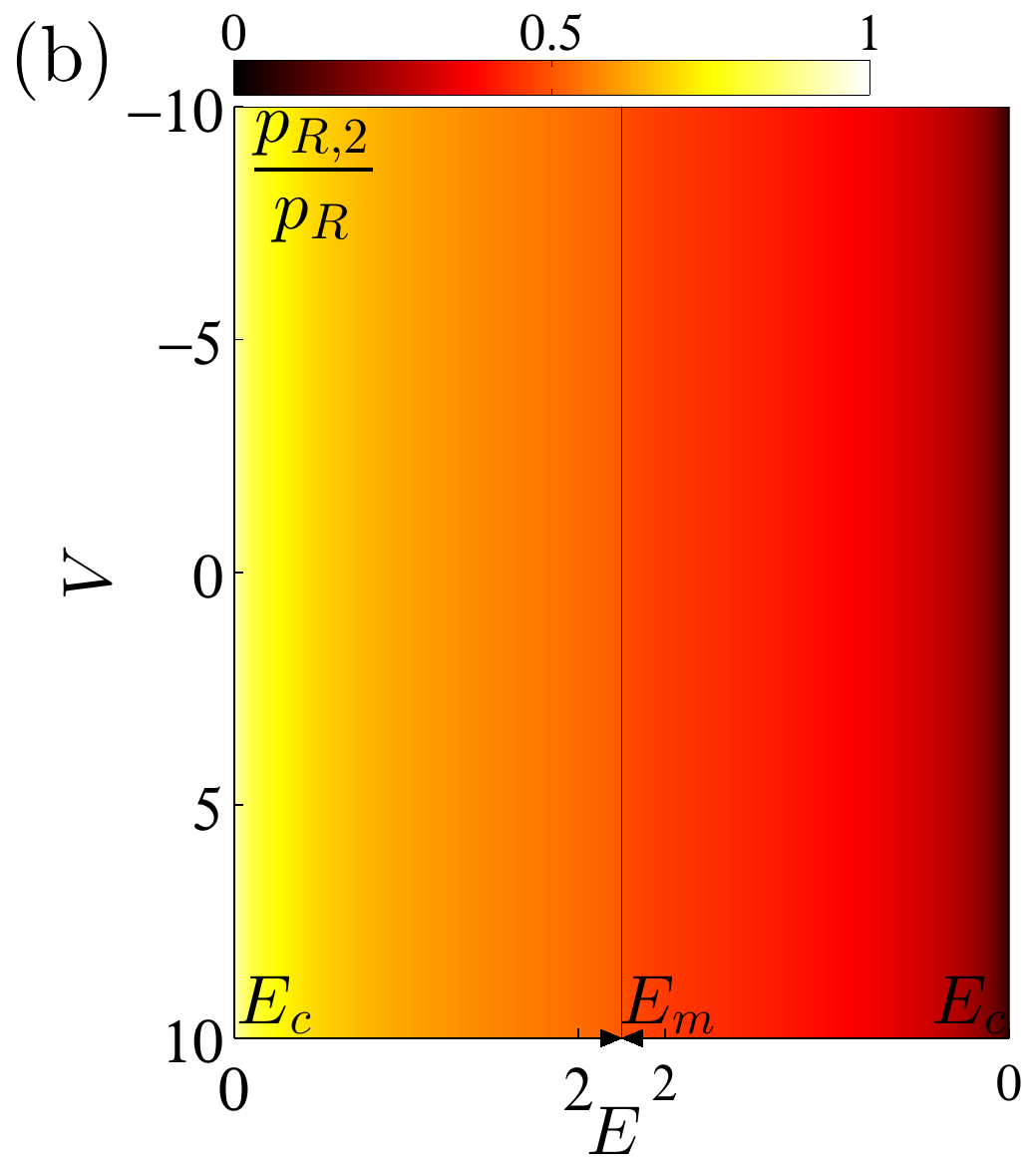}
\includegraphics[width=0.235\textwidth]{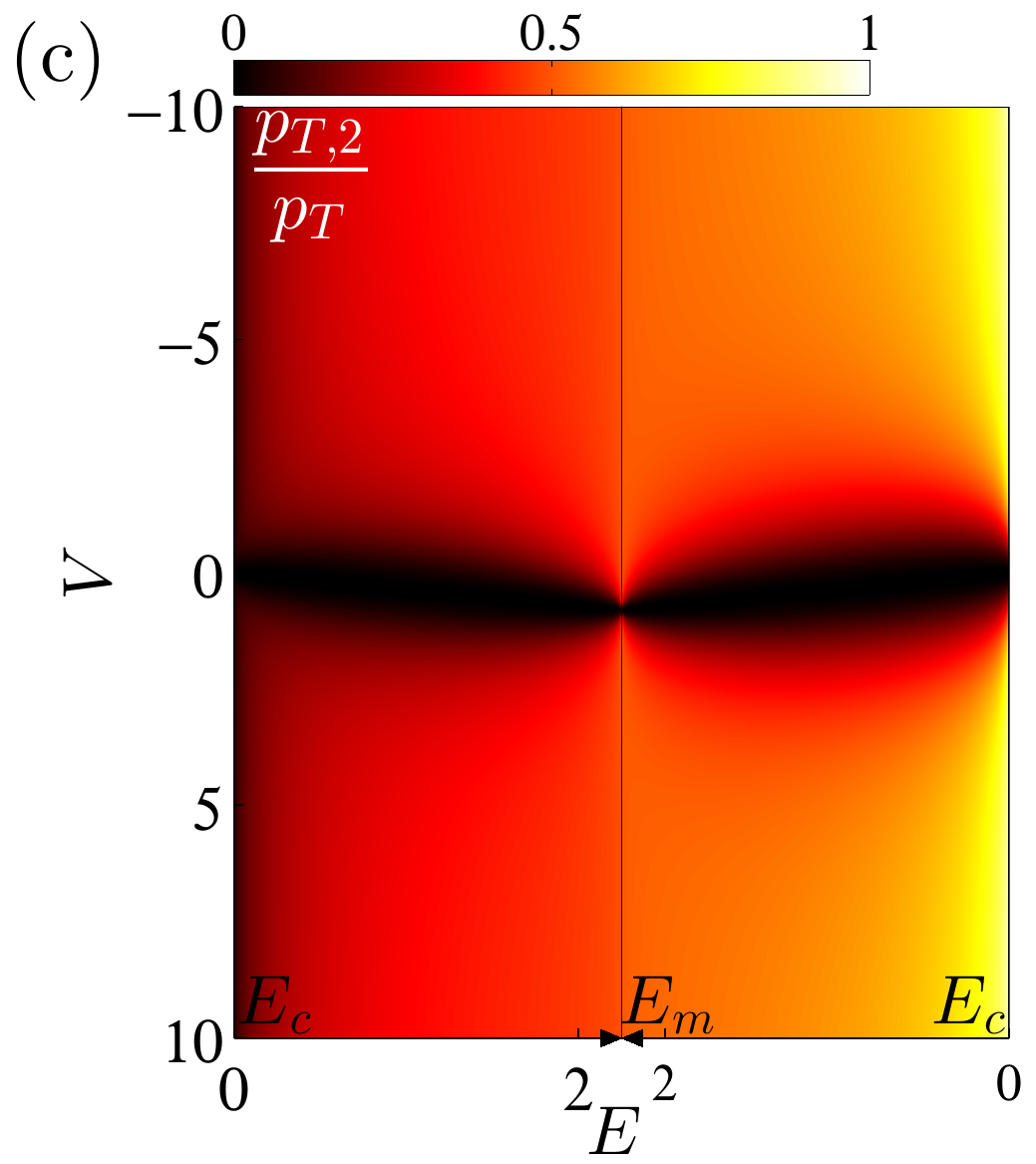}
\caption{(Color online) (a): Total transmission as a function of the incoming energy $E$ and the defect potential $V$.
(b,c) Branching ratios of reflection/transmission into the second channel to the total reflection/transmission.
Throughout, $t_1=t_2=1, \gamma=0.8$. Crosses mark the parameter values at which wave packet propagation runs are shown in Fig.~\ref{fig:prop3}.\label{fig:trans2}}
\end{figure}
Finally, let us have a look at the transmission and reflection probabilites as a function of the defect potential $V$ at a fixed coupling $\gamma$, as shown in Fig.~\ref{fig:trans2}.
A remarkable symmetry property of the scattering coefficients is found if $\gamma=1$ (no modulation of the hopping amplitudes due to the defect), in which case $V \rightarrow -V$ only causes complex conjugation of $T_{1,2}$, $R_{1,2}$ in the $E > E_c$ regime 
and thus the scattering probabilities become independent of the sign of the defect potential.
The remnants of this perfect symmetry at $\gamma=1$ can be seen in Fig.~\ref{fig:trans2}, which shows results for $\gamma=0.8$.
The second clear feature to be observed in Fig.~\ref{fig:trans2} is the region of large transmittivity at energies close to (and especially just below) $E_c$, independently of the defect potential $V$, as discussed above in Sec.~\ref{sec:sub}.
In Fig.~\ref{fig:prop3} we demonstrate that wide wave packets localized at momenta $k_1 \approx k_c$ indeed can pass the defect with little disturbance for vastly different values of $V$. 
The small deviations from perfect transmission we attribute to the inevitable finite width in momentum space, together with the fact that for $k_1 \gtrsim k_c$ the transmission probability rather quickly drops below unity.

\section{Summary and conclusions}
\label{sec:conclusions}
We have investigated the scattering problem in a discrete Schrödinger lattice with first- and second-neighbor hopping and a single-site defect, as can be realized with evanescently coupled optical waveguides or ultracold atoms.
Although the lattice is multiply connected, the continuity equation could be recast into the standard one-dimensional form by introducing a suitable generalized local current.
We explored bound states at the defect, noting in particular the absence of symmetry under the staggering transformation.
Turning to scattering solutions, an inspection of the band structure of the homogeneous system revealed two different energetic regimes,
separated by a critical energy marked by a transmission resonance, independently of the details of the defect.
At subcritical energies, one closed and one open channel coexist and give rise to Fano-Feshbach resonance phenomena at weak coupling, where analytical approximations to the Fano lineshape parameters were given
that demonstrated a large degree of tunability of the resonance asymmetry. 
At supercritical energies, there are two open channels which are coupled in the scattering process, giving rise to peculiar wave packet dynamics, where an incoming wave packet splits into multiple fragments moving at different group velocities.
The changes of the branching ratios of this splitting process when varying the model parameters were analyzed.\\
In several aspects, our results are reminiscent of those found for a cubic two-leg ladder with two transverse modes, where longitudinal and transverse motion asymptotically decouple \cite{Mizes1991}. 
That system, when perturbed by an immersed defect, for instance also exhibits a defect-independent transmission resonance
at the threshold energy for the opening of the second channel, and the analog of the wave packet splitting we observe would be a partial transfer into the other transverse mode during scattering,
which redistributes energy from the longitudinal motion (thus providing an intuitive explanation for the different group velocities of the outgoing wave packets).
Our model is fundamentally different from this cubic ladder because it does not admit the notion of separated longitudinal and transverse degrees of freedom, yet, to some extent the zigzag lattice 
may be thought of as a continuous deformation of the ladder, cf. \cite{Sukhorukov2008}. The increased connectivity of the one-dimensional chain due to the second-neighbor hopping to some 
extent mimicks a second (transverse) dimension.\\
We have chosen here the second-neighbor hopping model with the single-site defect because it is the most simple framework to observe the demonstrated effects and at the same time admits a transparent analytical treatment.
Extensions to other types of defects are straightforward, and also a more abstract access in terms of Green's functions has been sketched before \cite{Koiller1981,Schwalm1988}.
Using nanofiber-based optical traps of helix-shape for ultracold atoms \cite{Reitz2012},
realizations of related discrete models with sizable hopping to selected remote neighbors (also beyond the second one) may become accessible \cite{Stockhofe2015}, 
with three-dimensional helix arrangements of lattice sites generalizing the zigzag.
In that case, the band structure may exhibit multiple extrema within the first Brillouin zone and more channels are added to the problem,
suggesting the existence of multiple scattering resonances and enhanced wave packet fragmentation in scattering events.
\begin{acknowledgments}
We thank C. Morfonios for insightful discussions. J. S. gratefully acknowledges support from the Studienstiftung des deutschen Volkes.
\end{acknowledgments}

 \bibliography{bibliography}{}

\end{document}